\documentclass{WileyMSP-template}

\begin{document}

\pagestyle{fancy}
\rhead{\includegraphics[width=2.5cm]{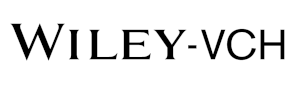}}


\newcommand{\del}{\partial}
\newcommand{\bs}{\boldsymbol}
\newcommand{\rmd}{\mathrm{d}}
\newcommand{\rmi}{\mathrm{i}}
\newcommand{\rminc}{\mathrm{inc}}
\newcommand{\rmsc}{\mathrm{sc}}
\newcommand{\rmtot}{\mathrm{tot}}
\newcommand{\rmtr}{\mathrm{tr}}
\newcommand{\rmin}{\mathrm{in}}
\newcommand{\rmout}{\mathrm{out}}
\newcommand{\rmoi}{\mathrm{oi}}
\newcommand{\rmio}{\mathrm{io}}
\newcommand{\rmEd}{\mathrm{Ed}}
\newcommand{\rmHd}{\mathrm{Hd}}
\newcommand{\rmkin}{\mathrm{kin}}
\newcommand{\rmkout}{\mathrm{kout}}

\title{A non-singular, field-only surface integral method for interactions between electric and magnetic dipoles and nano-structures}

\maketitle


\author{Qiang Sun*}
\author{Evert Klaseboer}


\dedication{}

\begin{affiliations}
Dr. Qiang Sun\\
Australian Research Council Centre of Excellence for Nanoscale BioPhotonics, School of Science, RMIT University, Melbourne, VIC 3001, Australia\\
Email Address: qiang.sun@rmit.edu.au

Dr. Evert Klaseboer\\
Institute of High Performance Computing, 1 Fusionopolis Way, Singapore, 138632, Singapore

\end{affiliations}


\keywords{Near-surface fields, Circularly polarized dipole}

\begin{abstract}

With the development of condensed-matter physics and nanotechnology, attention has turned to the fields near and on surfaces that result from interactions between electric dipole radiation and mesoscale structures. It is hoped that studying these fields will further our understanding of optical phenomena in nano-optics, quantum mechanics, electromagnetics and sensing using solid-state photon emitters. Here, we describe a method for implementing dynamic electric and magnetic dipoles in the frequency domain into a non-singular field-only surface integral method. We show that the effect of dipoles can conveniently be described as a relatively simple term in the integral equations, which fully represents how they drive the fields and interactions. Also, due to the non-singularity, our method can calculate the electric and magnetic fields on the surfaces of objects in both near and far fields with the same accuracy, which makes it an ideal tool to investigate nano-optical phenomena. The derivation of the framework is given and tested against a Mie theory alike formula. Some interesting examples are shown involving the interaction of dipoles with different types of mesoscale structures including parabolic nano-antennas and gold probes.

\end{abstract}


\section{Introduction}\label{sec:introduction}

The interaction of the radiation originating from dynamic electric dipoles with matter has a long history which can be traced back to the celebrated work of Lord Rayleigh~\cite{Strutt1871} (see also the classic book of van de Hulst~\cite{hulst1981light}) and can lead to interesting phenomena. Some communication technologies were developed based on such phenomena~\cite{King1992}, mainly using the far field pattern of dipole radiation since there was essentially little need to investigate near field phenomena. Recently however, with the development of condensed-matter physics and nanotechnology, the near fields / surface fields of interactions between dynamic electromagnetic dipole radiation and mesoscale structures have become the center of interest~\cite{Novotny2006}. A better understanding from a fundamental physics point of view can help to design better applications, such as in biophysics, novel optical phenomena in nano-optics~\cite{RodriguezFortuno2013, Bartolo2017, Picardi2017, Picardi2018, Picardi2019, VzquezLozano2019}, quantum mechanics electromagnetics~\cite{Keller2011}, imaging~\cite{Enderlein2006} and sensing~\cite{Antosiewicz2016} using solid-state photon emitters~\cite{Aharonovich2016} (for instance the defects in diamonds). As such, it is desirable to have at our disposal a robust method to effectively and efficiently calculate the electromagnetic fields driven by the dipole radiation and its interaction with the structure's surface, both in the near and in the far field. 

Due to the limited amount of available analytical solutions, even for simple geometric shapes, the interactions between dynamic electric dipoles and matter need to be calculated using numerical methods. From a computational point of view, numerical methods can be divided into roughly two classes; volume based methods and surface based methods. Almost all the differential equation-based methods, such as the finite element method (FEM)~\cite{Jin2014}, the finite-difference time-domain method (FDTD) \cite{KarimpourAnnalen2020} and the discrete-dipole approximation (DDA) \cite{KimAnnalen2020} belong to the volume based method class. Some integral equation-based methods such as the volumetric Method of Moments algorithm (V-MoM) \cite{HeOpticsExpress2018} and the (discontinuous) Galerkin time-domain~\cite{Cockburn2000} can also be regarded as belonging to the volume based method class. In order to use a volume based method to study the interaction between dipoles and objects, the complete simulation domain which includes both the scatterers and the surrounding medium needs to be discretized, which thus introduces significant computational overhead. Also, special attention is needed to handle the mesh in the domain around a dipole and how the mesh evolves to the surrounding space~\cite{Kalayeh2015}. Surface based methods, such as the method of moments (MoM)~\cite{Gallinet2015} and the $\phi-\bs{A}$ potential boundary element method \cite{GarciadeAbajo}, usually only require the discretization of the scattered surfaces. The most popular surface based method is MoM which was traditionally used to calculate the electromagnetic field of an electric dipole antenna for applications in Through-the-Earth communication. MoM can calculate the far field, but in spite of its popularity, performs much less accurate when accessing the near field or the surface field due to the congenital hypersingularies of this method. 
The $\phi-\bs{A}$ potentials boundary element method solves for the scalar and vector potentials $\phi$ and $\boldsymbol A$. In order to get back the electric field, one must take a further step to calculate the electric field by differentiation, 
which might reduce the accuracy of the calculation. 
So we are left with the undesirable situation that both the popular volume-based and surface-based methods are not ideally suited to study the interactions between the radiation of dynamic electric dipoles and objects in mesoscale for nano-optics and an alternative method which can deliver more accurate results would be welcomed.

Recently, a new type of surface integral method, the fully non-singular field-only surface method has been developed by us to study the interactions between matter and an incoming electromagnetic wave~\cite{Klaseboer2017, Sun2017, Sun2020a, Sun2020b, Sun2021}. In this paper, we develop a method to include electric dipoles into this framework. Before we step into the details, we  briefly summarise the advantages of our method: 
\begin{itemize}
  \item The electric dipole is implemented explicitly on the right hand side of the numerical system, which is derived analytically. From a physics point of view, such a system fully represents how dynamic dipoles drive the fields and interactions.  
  \item Our method only needs a mesh on the boundary of the scatterer. For volume methods, particular attention on meshing near the dipole is needed. It is even more difficult to handle multiple dipoles, while in our implementation this is straightforward. 
  \item The surface integration method automatically takes care of radiation conditions at infinity.
  \item Compared to other surface integral methods, our method has two obvious advantages. Firstly, our method solves the fields of physical interest directly, namely the electric and magnetic fields. However, other surface integral methods solve for some intermediate quantities, such as surface currents or potentials~\cite{Hohenester2012,GarciadeAbajo}, while the electric and magnetic fields are obtained by post-processing in which the accuracy might be compromised. Secondly, there are no singularities in the surface integrals over the scatterers due to the Green's function and its derivative. Consequently, high order surface elements, such as quadratic triangular elements, can be easily employed to obtain highly accurate results with fewer degrees of unknowns. 
  \item As our method is fully free of singularities, we can get the electric and magnetic fields on the surfaces of objects, in both near and far fields with the same accuracy, which makes our method an ideal tool to investigate nano-optical phenomena.
\end{itemize}

The paper is outlined as follows. In the next section, the description of how to include the electric dipole in the numerical framework is given, followed by the validation using the Mie theory. Then some examples for nano-optics are shown, and after a discussion section on magnetic dipoles, the paper is terminated with the conclusions.

\section{Non-singular field-only surface integral method}

To clearly illustrate our non-singular field-only surface method for the interactions between dynamic electromagnetic dipoles and mesoscale objects, we start with a simple case of an electric dipole near a perfect electric conductor. Then, we extend the framework for the interaction between an electric dipole and a dielectric object. The same idea can be further extended to multiple electric and magnetic dipoles and multiple scatterers, as discussed in Sec.~\ref{sec:discussion}.

\subsection{Electric dipole near a perfect electric conductor}\label{sec:pec}

 In the frequency domain, the electric and magnetic field due to an electric dipole are, respectively
 , 
\begin{subequations}\label{eq:EH_Edipole}
\begin{align}
    \bs{E}^{\rmEd} & = \frac{1}{4\pi \epsilon_0 \epsilon_r } \nabla \times \left \{ \nabla \left[\frac{\exp( \rmi k |\bs{x} - \bs{x}_{d}|)}{|\bs{x} - \bs{x}_{d}|}\right] \times \bs{p} \right\}, \label{eq:E_Edipole}\\
    \bs{H}^{\rmEd} & = \frac{\omega}{4\pi \rmi} \left \{ \nabla \left[\frac{\exp( \rmi k |\bs{x} - \bs{x}_{d}|)}{|\bs{x} - \bs{x}_{d}|}\right] \times \bs{p} \right\} \label{eq:H_Edipole}
\end{align}
\end{subequations}
where $\bs{p}$ is the electric dipole momentum vector (with dimension Coulomb meter), $\bs{x}$ is the observation location and $\bs{x}_{d}$ the location of the dipole (presumed to be fixed in space, to avoid Doppler effects), $k = \omega /c$ is the wavenumber with $\omega$ the angular frequency and $c$ the speed of light, $\epsilon_0$ and $\epsilon_r$ are the permittivity of free space and relative permittivity of the medium respectively. 

\begin{figure}[t]
\centering
\includegraphics[width=0.5\textwidth]{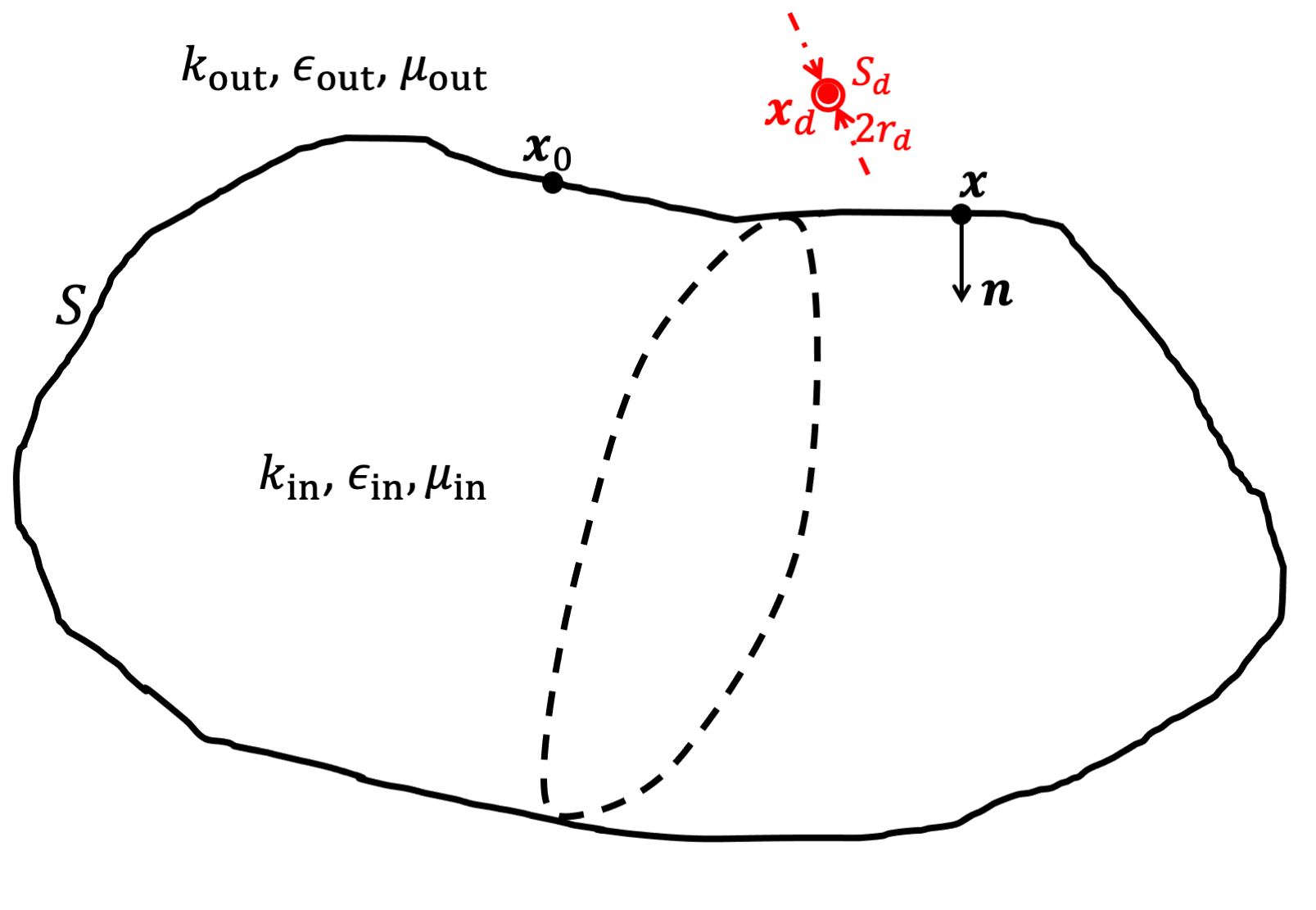}
\caption{Object with surface $S$ embedded in an infinite outer domain with parameters $k_{\rmout}$, $\epsilon_{\rmout}$, $\mu_{\rmout}$ (the wavenumber, permittivity and permeability respectively). An electric dipole is situated at the position $\bs x_d$. The normal vector $\bs n$ is pointing into the object. For a dielectric object we also take into account the inner parameters $k_{\rmin}$, $\epsilon_{\rmin}$, $\mu_{\rmin}$. In the theory a vanishing small volume around the dipole with surface $S_d$ and radius $r_d$ is excluded.} \label{Fig:1sketch}
\end{figure}

In the domain that excludes the dipole, thus the domain outside the scatterer $S$ and a tiny spherical surface $S_{d}$ that encloses the dipole as shown in Fig.~\ref{Fig:1sketch}, the Maxwell equations in the frequency domain reduce to the following two equations for the electric field $\bs E$
\begin{subequations}\label{eq:MaxwellEq2}
    \begin{align}
        &\nabla^2 \bs{E} + k^2 \bs{E} = \bs{0}, \label{eq:HelmE} \\
        &\nabla \cdot \bs{E} = 0. \label{eq:divE2} 
    \end{align}
\end{subequations}
Thus the electric field satisfies a vector Helmholtz equation and is simultaneously divergence free. 
To obtain the interactions between one or more dipoles and scatterers in a homogeneous medium, it is necessary to efficiently and accurately solve Eq.~(\ref{eq:MaxwellEq2}). 

It has been demonstrated that Eq. (\ref{eq:HelmE}) can be solved effectively and accurately by the recently developed non-singular field only surface integral method~\cite{Sun2020a, Sun2020b}. To obtain the electric field on the boundary, each Cartesian component of the $\bs{E}$ field in Eq.~(\ref{eq:HelmE}), which obeys the scalar Helmholtz equation (an elliptic partial differential equation), is represented by the non-singular boundary integral equation~\cite{Sun2021b} as (see Appendix~\ref{app:integral} for derivation)
\begin{equation} \label{eq:nsbim}
    \begin{aligned}
4\pi \bs{E}(\boldsymbol x_0) + \int_{S}  \left[\bs{E}(\boldsymbol x)  H_k(\bs{x},\bs{x}_0) -\bs F (\boldsymbol x)  H_0(\bs{x},\bs{x}_0)\right] \rmd S(\bs{x}) \\
= \int_{S} \left[\frac{\del \bs{E}(\boldsymbol x) } {\del n}  G_k(\bs{x},\bs{x}_0) - \frac{\del \bs F (\boldsymbol x) } {\del n}  G_0(\bs{x},\bs{x}_0) \right] \rmd S(\bs{x}) - \bs{I}^{\rmEd}.
\end{aligned}
\end{equation}
where $\bs x_0$ is an observation point and $\bs x$ is an integration point, as shown in Fig.~\ref{Fig:1sketch}. $G_k(\bs x,\bs x_0) = \exp(\rmi k|\bs x - \bs x_0|)/|\bs x - \bs x_0|$ is the Green's function for the Helmholtz equation and $G_0=1/|\bs x - \bs x_0|$ the Laplace equivalent. $H_k = \partial G_k/\partial n$ is the normal derivative of the Green's function as in the standard boundary element framework with $\partial/\partial n = \bs n \cdot \nabla$ where $\bs{n}$ is the unit normal vector pointing into the object. The classical boundary integral equation is made non-singular analytically by subtracting a function $\bs F(\bs x)$:
\begin{equation}\label{eq:nsbim_psiE}
\begin{aligned}
    \bs F(\boldsymbol x) = \bs{E}(\boldsymbol x_0) + \boldsymbol n(\boldsymbol x_0) \cdot (\boldsymbol x - \boldsymbol x_0) \frac{\del \bs{E}(\boldsymbol x_0)}{\del n}.
\end{aligned}
\end{equation}
which satisfies the Laplace equation as $\nabla^2 \bs F(\boldsymbol x) = \bs 0 $. Note that both terms in each bracket of Eq.~(\ref{eq:nsbim}) cancel each other out when $\bs x$ approaches $\bs x_0$ and effectively eliminate the singularities analytically which arise due to the Green's function. 

The term $\bs I^{\rmEd}$ represents the integral on $S_{d}$ which is a tiny spherical surface with vanishing radius $r_d$ that encloses the dipole:
\begin{equation}\label{eq:nsbim_Id}
\begin{aligned}
    \bs{I}^{\rmEd} = \lim_{r_d \rightarrow 0}\int_{S_{d}} \left[ H_{k}(\bs{x},\bs{x}_0)  
    \bs{E}(\bs{x}) - G_{k}(\bs{x},\bs{x}_0) \frac{\del \bs{E}(\bs{x})}{\del n}   \right] \rmd S(\bs{x}).
\end{aligned}
\end{equation}
In Appendix~\ref{app:integral}, it is shown that the term $\bs I^{\rmEd}$ can be written as: 
\begin{equation} \label{eq:Iterm}
\begin{aligned}
    \bs{I}^{\rmEd} &=
       -\frac{1}{\epsilon_0 \epsilon_r } \nabla_d \times \left \{ \nabla_d  G_{k}(\bs{x}_d,\bs{x}_0) \times \bs{p} \right\} = -\frac{1}{\epsilon_0 \epsilon_r } \nabla \times \left \{ \nabla  G_{k}(\bs{x}_0,\bs{x}_d) \times \bs{p} \right\} \\
       &=-\frac{\exp(\rmi k r_{0d})}{\epsilon_0 \epsilon_r r_{0d}^3}\left\{(-k^2 r_{0d}^2 - 3 \rmi k r_{0d} + 3) \frac{(\bs x_{0d}\cdot \bs p)}{r_{0d}^2} \bs x_{0d}  +(k^2 r_{0d}^2 + \rmi k r_{0d}-1)\bs p  \right\}
\end{aligned}
\end{equation}
in which $\bs x_{0d} = \bs x_0 - \bs x_d$, $r_{0d}=|\bs x_{0d}|$ and $G_{k}(\bs{x}_d,\bs{x}_0) = G_{k}(\bs{x}_0,\bs{x}_d)$ are used. 

The above boundary integral equation~(\ref{eq:nsbim}) gives a relationship between the Cartesian components of the electric field ($E_x$, $E_y$, $E_z$) and their normal derivatives ($\partial E_x/\partial n$, $\partial E_y/\partial n$, $\partial E_z/\partial n$) on the surface $S$. However, the divergence free condition in the domain, $\nabla \cdot \bs E = 0$, also needs to be satisfied simultaneously. Sun et al.~\cite{Sun2020a} demonstrated that when the divergence free condition is satisfied everywhere on the surface of the object, it is automatically satisfied everywhere in the domain. It has also been shown in Sun et al.~\cite{Sun2020a} that the divergence free condition $\nabla \cdot \bs E = 0$ is equal to
\begin{equation} \label{eq:divE0}
    \bs n \cdot \frac{\partial \bs E}{\partial n} - \kappa E_n + \frac{\partial E_{t_1}}{\partial t_1} + \frac{\partial E_{t2}}{\partial t_2} = 0.
\end{equation}
Here $\kappa$ is the curvature on the surface, $E_n \equiv \bs{E} \cdot \bs{n}$ is the normal component of the electric field on the surface, and $E_{t1}\equiv \bs{E} \cdot \bs{t}_1$, $E_{t2}\equiv \bs{E} \cdot \bs{t}_2$ are the tangential components with $\bs{t}_1$ and $\bs{t}_2$ the two orthogonal surface unit tangential vectors and $\partial/\partial t_1 = \boldsymbol t_1 \cdot \nabla$ (and similar for $\partial/\partial t_2$). 

Since the boundary conditions are given in terms of tangential components of the electric field, it turns out to be easier to use the normal and tangential components as unknowns in our framework. As such, we rewrite the electric field and its normal derivative as
\begin{equation}
    \begin{aligned} \label{eq:Vdecomposition}
        \bs E &= E_x \bs e_x + E_y \bs e_y + E_z \bs e_z 
         = E_n \bs n + E_{t1} \bs t_1 + E_{t2} 
        \bs t_2,    \\
         \frac{\partial \bs E}{\partial n} &= \frac{\partial E_x}{\partial n} \bs e_x + \frac{\partial E_y}{\partial n} \bs e_y + \frac{\partial E_z}{\partial n} \bs e_z = \left(\bs n \cdot \frac{\partial \bs E}{\partial n}\right) \bs n + \left(\bs t_1 \cdot \frac{\partial \bs E}{\partial n}\right)  \bs t_1
         +\left(\bs t_2 \cdot \frac{\partial \bs E}{\partial n}\right) \bs t_2.
         \end{aligned}
\end{equation}
Thus for example $E_x$ can be expressed as $E_x = E_n (\bs n \cdot \bs e_x) + E_{t1} (\bs t_1 \cdot \bs e_x) + E_{t2} (\bs t_2 \cdot \bs e_x)  = E_n n_x + E_{t1} t_{1x} + E_{t2} t_{2x}$.

To numerically solve Eq.~(\ref{eq:nsbim}) for the interaction between a dynamic electric dipole and a PEC object, the object surface is discretised with $N$ nodes. Then, the left hand side of Eq.~(\ref{eq:nsbim}) results in a matrix equation relating all $x$-components of the electric field as: ${\cal{H}} E_x$ and its normal derivatives on the right hand side of Eq.~(\ref{eq:nsbim}) as: ${\cal{G}} \partial E_x/ \partial n$. Here, ${\cal{G}}$ and ${\cal{H}}$ are $N \times N$ matrices and $E_x$ and $\partial E_x/\partial n$ are now column vectors of size $N$. A similar procedure applies for the $y$ and $z$-components. To introduce boundary condition in Eq.~(\ref{eq:divE0}) into the linear system, the Cartesian components of the electric field and its normal derivatives can be reconstructed from the normal and tangential components with Eq.~\ref{eq:Vdecomposition}. Also, for a PEC object, the surface tangential components of the electric field are zeros. As such, we get the following $3N \times 3N$ matrix system as
  \label{eq:linear_system}
  \begin{equation}
    \label{eq:linear_system_part_a}
    \begin{bmatrix}
      n_{x} ({\cal{H}} - \kappa {\cal{G}}) & -t_{1x} {\cal{G}} & -t_{2x} {\cal{G}} \\
      n_{y} ({\cal{H}} - \kappa {\cal{G}}) & -t_{1y} {\cal{G}} & -t_{2y} {\cal{G}} \\
      n_{z} ({\cal{H}} - \kappa {\cal{G}}) & -t_{1z} {\cal{G}} & -t_{2z} {\cal{G}}
    \end{bmatrix}
    \begin{bmatrix}
      E_n^\rmsc \\ \bs{t}_{1} \cdot \frac{\partial \bs{E}^\rmsc}{\partial n} \\ \bs{t}_{2} \cdot \frac{\partial \bs{E}^\rmsc}{\partial n}
    \end{bmatrix}
    =
    \begin{bmatrix}
      -I_{x}^{\rmEd}\\ -I_{y}^{\rmEd} \\ -I_{z}^{\rmEd} 
    \end{bmatrix},
  \end{equation} 
from which the electric field on the PEC surface can be obtained. The interested reader can find more details on the implementation in Sun et al.~\cite{Sun2020a}.

\subsection{Electric dipole in and/or near a dielectric object}~\label{sec:dielectrics}

For a dielectric object, we need to calculate the field inside the object $\bs E^{\rmin}$ as well as the external electric field on both sides of the surface $\bs E^{\rmout}$. The internal domain is characterised by $k_{in}$, $\epsilon_{\rmin}$, $\mu_{\rmin}$ while the outer domain by $k_{\rmout}$,  $\epsilon_{\rmout}$ and $\mu_{\rmout}$. The field only surface integral equation for the external domain s:
\begin{equation} \label{eq:nsbimOUT2}
    \begin{aligned}
4\pi \bs{E}^{\rmout}(\boldsymbol x_0) + \int_{S}  \left[\bs{E}^{\rmout}(\boldsymbol x)  H_{\rmkout}(\bs{x},\bs{x}_0) -\bs F^{\rmout} (\boldsymbol x)  H_0(\bs{x},\bs{x}_0)\right] \rmd S(\bs{x}) \\
= \int_{S} \left[\frac{\del \bs{E}^{\rmout}(\boldsymbol x) } {\del n}  G_{\rmkout}(\bs{x},\bs{x}_0) - \frac{\del \bs F^{\rmout} (\boldsymbol x) } {\del n}  G_0(\bs{x},\bs{x}_0) \right] \rmd S(\bs{x}) - \bs{I}^{\rmEd}_{\rmout}.
\end{aligned}
\end{equation}
For the internal domain we find: 
\begin{equation} \label{eq:nsbimIN2}
\begin{aligned}
 &\int_{S}  \left[\bs{E}^{\rmin}(\boldsymbol x)  H_{\rmkin}(\bs{x},\bs{x}_0) -\bs F^{\rmin} (\boldsymbol x)  H_0(\bs{x},\bs{x}_0)\right] \rmd S(\bs{x}) \\
=& \int_{S} \left[\frac{\del \bs{E}^{\rmin}(\boldsymbol x) } {\del n}  G_{\rmkin}(\bs{x},\bs{x}_0) - \frac{\del \bs F^{\rmin} (\boldsymbol x) } {\del n}  G_0(\bs{x},\bs{x}_0) \right] \rmd S(\bs{x}) - \bs{I}^{\rmEd}_{\rmin}.
\end{aligned}
\end{equation}
In Eqs.~(\ref{eq:nsbimOUT2}) and (\ref{eq:nsbimIN2}), $\bs F^{\rmout} (\boldsymbol x) = \bs{E}^{\rmout}(\boldsymbol x_0) + \boldsymbol n(\boldsymbol x_0) \cdot (\boldsymbol x - \boldsymbol x_0) \frac{\del \bs{E}^{\rmout}(\boldsymbol x_0)}{\del n}$ and similar for $\bs F^{in}$ as $\bs F^{\rmin} (\boldsymbol x) = \bs{E}^{\rmin}(\boldsymbol x_0) + \boldsymbol n(\boldsymbol x_0) \cdot (\boldsymbol x - \boldsymbol x_0) \frac{\del \bs{E}^{\rmin}(\boldsymbol x_0)}{\del n}$. Note the difference of a term with $4\pi \boldsymbol E^{\rmout}$ between Eqs.~(\ref{eq:nsbimOUT2}) and (\ref{eq:nsbimIN2}) due to the absence of integrals at infinity for the internal problem in Eq.~(\ref{eq:nsbimIN2}) (see Appendix~\ref{app:integral}). 

The boundary conditions for the electric field, expressed in terms of the normal and tangential components, are:
\begin{subequations} \label{eq:BC}
\begin{align}
    E_n^{\rmin}&=\epsilon_{\rmoi} E_n^{\rmout} \label{eq:BCEn}\\
    E_{t1}^{\rmin}&=E_{t1}^{\rmout}\label{eq:BCEt1}\\
    E_{t2}^{\rmin}&=E_{t2}^{\rmout} \label{eq:BCEt2}
    \end{align}
\end{subequations}
with $\epsilon_{\rmoi}=\epsilon_{\rmout}/\epsilon_{\rmin}$ the ratio of permittivities between the two media. Meanwhile, the divergence condition of Eq.~(\ref{eq:divE0}) is applied to $\bs E^{\rmout}$ as $\bs n \cdot \frac{\partial \bs E^{\rmout}}{\partial n} - \kappa E_n^{\rmout} + \frac{\partial E_{t_1}^{\rmout}}{\partial t_1} + \frac{\partial E_{t2}^{\rmout}}{\partial t_2} = 0$ and to $\bs E^{\rmin}$ as $\bs n \cdot \frac{\partial \bs E^{\rmin}}{\partial n} - \kappa E_n^{\rmin} + \frac{\partial E_{t_1}^{\rmin}}{\partial t_1} + \frac{\partial E_{t2}^{\min}}{\partial t_2} = 0$. In order to get rid of the tangential derivative terms these two equations are subtracted while the boundary conditions of Eq.~(\ref{eq:BCEt1}) and (\ref{eq:BCEt2}) are used. As such, we can express $E_n^{\rmin}$ in terms of $E_n^{\rmout}$ with Eq.~(\ref{eq:BCEn}):
\begin{equation} \label{eq:ndEindn}
    \bs n \cdot \frac{\partial \bs E^{\rmin}}{\partial n}=\kappa\left(\epsilon_{\rmoi}-1 \right) E_n^{\rmout} + \bs n \cdot \frac{\partial \bs E^{\rmout}}{\partial n}.
\end{equation}
This equation relates the normal component of the normal derivatives,  $\bs n \cdot (\partial \bs E/\partial n)$, at both sides of the boundary of the object. We still need another equation to relate both tangential components of the normal derivatives: $\bs t \cdot (\partial \bs E/\partial n)$. This can be done by using the tangential continuity of the magnetic field on the dielectric surface: $H_{t1}^{\rmin} = H_{t1}^{\rmout}$ and $H_{t2}^{\rmin} = H_{t2}^{\rmout}$. Using the convention $\bs n = \bs t_1 \times \bs t_2$, $\bs t_1 \cdot \bs t_2 = 0$, we express $H_{t1}$ and $H_{t2}$ in terms of the electric field on the surface as
\begin{equation}
    \begin{aligned}
H_{t1}= \bs t_1 \cdot \bs H =\bs t_2 \cdot (\bs n \times \bs H) = \frac{1}{\rmi \omega \mu}\bs t_2 \cdot (\bs n \times \nabla \times \bs E)=\frac{1}{\rmi \omega \mu}\left[\bs t_1 \cdot \frac{\partial \bs E}{\partial n} - \bs n \cdot \frac{\partial \bs E}{\partial t_1} \right],\\
-H_{t2}= -\bs t_2 \cdot \bs H = \bs t_1 \cdot (\bs n \times \bs H) =\frac{1}{\rmi \omega \mu}\bs t_1 \cdot (\bs n \times \nabla \times \bs E)=\frac{1}{\rmi \omega \mu}\left[\bs t_2 \cdot \frac{\partial \bs E}{\partial n} - \bs n \cdot \frac{\partial \bs E}{\partial t_2} \right].
  \end{aligned}
\end{equation} 
It is worth noting the inversion of $t_1$ and $t_2$ and the appearance of a minus sign in the $H_{t2}$ term in the above equation. 

For simplicity, we assume that $\mu_{\rmin}=\mu_{\rmout}$ (if this is not the case, see Sun et al.~\cite{Sun2020b}), and then the tangential continuity of $\bs H$ across the dielectric surface is identical to
\begin{equation}
\begin{aligned} \label{eq:Htangcond}
    \bs t_1 \cdot \frac{\partial \bs E^{\rmin}}{\partial n} - \bs n \cdot \frac{\partial \bs E^{\rmin}}{\partial t_1} = \bs t_1 \cdot \frac{\partial \bs E^{\rmout}}{\partial n} - \bs n \cdot \frac{\partial \bs E^{\rmout}}{\partial t_1}, \\
    \bs t_2 \cdot \frac{\partial \bs E^{\rmin}}{\partial n} - \bs n \cdot \frac{\partial \bs E^{\rmin}}{\partial t_2} = \bs t_2 \cdot \frac{\partial \bs E^{\rmout}}{\partial n} - \bs n \cdot \frac{\partial \bs E^{\rmout}}{\partial t_2}.
    \end{aligned}
\end{equation}
The terms with $\bs n \cdot \partial \bs E/ \partial t_1$ and $\bs n \cdot \partial \bs E/ \partial t_2$ can be rewritten as\footnote{This can easiest be seen by writing the vector $\bs E$ into normal and tangential components as: $\partial \bs E/\partial t_1 = \frac{\partial}{\partial t_1}[E_n \bs n + E_{t1} \bs t_1 + E_{t2} \bs t_2] = \bs n \frac {\partial E_n}{\partial t_1} + E_n \frac{\partial \bs n}{\partial t_1} + \bs t_1 \frac{\partial E_{t1}}{\partial t_1} + E_{t1} \frac{\partial \bs t_1}{\partial t_1} + \bs t_2 \frac{\partial E_{t2}}{\partial t_1} + E_{t2} \frac{\partial \bs t_2}{\partial t_1}$. Using $\partial \bs n/\partial t_1 = - \kappa_1 \bs t_1$, $\partial \bs t_1/ \partial t_1 = \kappa_1 \bs n$ and $\partial \bs t_2/ \partial t_1 = \bs 0$, we get $\bs n \cdot \frac {\partial \bs E}{\partial t_1} = \frac{\partial E_n}{\partial t_1} + \kappa_1 E_{t1}$ and similar for $\bs n \cdot \frac{\partial \bs E}{\partial t_2}= \frac{\partial E_n}{\partial t_2} + \kappa_2 E_{t2}$. 
}  $\bs n \cdot \partial \bs E/ \partial t_1 = \partial E_n/\partial t_1 +\kappa_1 E_{t1}$ and $\bs n \cdot \partial \bs E/ \partial t_2 = \partial E_n/\partial t_2 +\kappa_2 E_{t2}$, with $\kappa_1$ the curvature in the $\bs t_1$ direction and $\kappa_2$ the curvature in the $\bs t_2$ direction. Putting this into Eq.~(\ref{eq:Htangcond}) where the terms with $\kappa_1 E_{t1}$ and $\kappa_2 E_{t2}$ cancel out due to Eqs.~(\ref{eq:BCEt1}) and (\ref{eq:BCEt2}), and further using Eq.~(\ref{eq:BCEn}) to express $E_n^{\rmin}$ in terms of $E_n^{\rmout}$, we have $\frac{\partial E_n^{\rmout}}{\partial t_1} + \kappa_1 E_{t1}^{\rmout} - \frac{\partial E_n^{\rmin}}{\partial t_1} - \kappa_1 E_{t1}^{\rmin} = \frac{\partial E_n^{\rmout}}{\partial t_1} (1 - \epsilon_{\rmoi})$. As such,
\begin{equation} \label{eq:tdEindn}
\begin{aligned} 
    \bs t_1 \cdot \frac{\partial \bs E^{\rmin}}{\partial n} =\bs t_1 \cdot \frac{\partial \bs E^{\rmout}}{\partial n} + \frac{\partial E_n^{\rmout}}{\partial t_1} (\epsilon_{\rmoi}-1),\\
    \bs t_2 \cdot \frac{\partial \bs E^{\rmin}}{\partial n} =\bs t_2 \cdot \frac{\partial \bs E^{\rmout}}{\partial n} + \frac{\partial E_n^{\rmout}}{\partial t_2} (\epsilon_{\rmoi}-1).
    \end{aligned}
\end{equation}
With Eq.~(\ref{eq:BC}), we now can express all the components of $\bs E^{\rmin}$ in terms of $\bs E^{\rmout}$. All components of the normal derivatives of the internal field can also be expressed in terms of outer field variables with Eq.~(\ref{eq:ndEindn}) and Eq.~(\ref{eq:tdEindn}). The Cartesian $x$, $y$ and $z$ components can then be obtained with Eq.~(\ref{eq:Vdecomposition}). 

To numerically solve Eqs.~(\ref{eq:nsbimOUT2}) and (\ref{eq:nsbimIN2}) for the interaction between a dynamic electric dipole and a dielectric object, the object surface is discretised with $N$ nodes. Meanwhile, the $x$, $y$, $z$ components of $\bs E^{\rmin}$ and $\bs E^{\rmout}$ and their normal derivatives are expressed in terms of $E_n^{\rmout}$, $E_{t1}^{\rmout}$, $E_{t2}^{\rmout}$, $\bs n \cdot (\partial \bs E^{\rmout}/\partial n)$, $\bs t_1 \cdot (\partial \bs E^{\rmout}/\partial n)$ and $\bs t_2 \cdot (\partial \bs E^{\rmout}/\partial n)$, which results in a $6N \times 6N$ matrix system as
\begin{subequations}\label{eq:big_linear_system_full}
  \begin{equation}\label{eq:big_linear_system}
    \begin{bmatrix}
      n_{x}{\cal{H}}_{\rmout} & t_{1x}{\cal{H}}_{\rmout} & t_{2x}{\cal{H}}_{\rmout}
      & - {n_{x}}{\cal{G}}_{\rmout} & - {t_{1x}}{\cal{G}}_{\rmout}
      & - {t_{2x}}{\cal{G}}_{\rmout} \\
      n_{y}{\cal{H}}_{\rmout} & t_{1y}{\cal{H}}_{\rmout} & t_{2y}{\cal{H}}_{\rmout}
      & - {n_{y}}{\cal{G}}_{\rmout} & - {t_{1y}}{\cal{G}}_{\rmout}
      & - {t_{2y}}{\cal{G}}_{\rmout} \\
      n_{z}{\cal{H}}_{\rmout} & t_{1z}{\cal{H}}_{\rmout} & t_{2z}{\cal{H}}_{\rmout}
      & - {n_{z}}{\cal{G}}_{\rmout} & - {t_{1z}}{\cal{G}}_{\rmout}
      & - {t_{2z}}{\cal{G}_{\rmout}} \\
       \bar{{\cal{H}}}^{x}_{\rmin}  & t_{1x}{\cal{H}}_{\rmin}
      &   t_{2x}{\cal{H}}_{\rmin}  & -n_{x}{\cal{G}}_{\rmin}
      & - t_{1x} {\cal{G}}_{\rmin} & - t_{2x} {\cal{G}}_{\rmin}  \\
      \bar{{\cal{H}}}^{y}_{\rmin}  & t_{1y}{\cal{H}}_{\rmin}
      &   t_{2y}{\cal{H}}_{\rmin}  &  -n_{y}{\cal{G}}_{\rmin}
      & - t_{1y} {\cal{G}}_{\rmin} & - t_{2y} {\cal{G}}_{\rmin}  \\ 
      \bar{{\cal{H}}}^{z}_{\rmin}  & t_{1z}{\cal{H}}_{\rmin}
      &   t_{2z}{\cal{H}}_{\rmin}  &  -n_{z}{\cal{G}}_{\rmin}
      & - t_{1z} {\cal{G}}_{\rmin} & - t_{2z} {\cal{G}}_{\rmin}
    \end{bmatrix}
    \cdot
    \begin{bmatrix}
      E_n^{\rmout} \\ E_{t_{1}}^{\rmout} \\ E_{t_{2}}^{\rmout} \\
      \bs{n} \cdot \frac{\partial \bs{E}^{\rmout}}{\partial n}   \\
      \bs{t}_{1} \cdot \frac{\partial \bs{E}^{\rmout}}{\partial n}  \\
      \bs{t}_{2} \cdot \frac{\partial \bs{E}^{\rmout}}{\partial n}
    \end{bmatrix}
    =
    \begin{bmatrix}
      I^{\rmEd}_{\rmout,x} \\ 
      I^{\rmEd}_{\rmout,y} \\
      I^{\rmEd}_{\rmout,z} \\
      I^{\rmEd}_{\rmin,x} \\ 
      I^{\rmEd}_{\rmin,y} \\ 
      I^{\rmEd}_{\rmin,z}
    \end{bmatrix}, 
  \end{equation}
  where (with $\alpha = x,y,z.$)
  \begin{align}
    \bar{{\cal{H}}}^{\alpha}_{\rmin} &= \epsilon_{\rmoi} n_{\alpha} {\cal{H}}_{\rmin} + \left(\epsilon_{\rmoi}-1\right) {\cal{G}}_{\rmin} \left[
    - \kappa n_{\alpha} 
    - t_{1\alpha}\frac{\partial }{\partial t_1}
    -  t_{2\alpha} 
    \frac{\partial}{\partial t_2} \right], \label{eq:matrix_patial_derivative} 
  \end{align}
\end{subequations}
In the above matrix system, the first line corresponds to ${\cal{H}}_{\rmout} E_x^{\rmout} = {\cal{G}}_{\rmout} \partial E_x^{\rmout}/\partial n$, the matrix equivalent of Eq.~(\ref{eq:nsbimOUT2}) for the $x$-component. Line 2 and 3 correspond to the $y$ and $z$-component, respectively. Lines 4, 5 and 6 correspond to the matrix equivalent of Eq.~(\ref{eq:nsbimIN2}) for the $x$, $y$ and $z$-component of $\boldsymbol E^{\rmin}$, respectively. The implementation is straightforward, expect perhaps for the tangential derivative matrices $\partial/\partial t_1$ and $\partial/ \partial t_2$ (for more details see Sun et al.~\cite{Sun2020b}). Once the matrix system is solved we can get back the Cartesian components with Eq.~(\ref{eq:Vdecomposition}). 

As we now have the electric field and its normal derivative on the surface for both the internal and external domain, we can get the electric field anywhere in the domain by postprocessing (see Sun et al.~\cite{Sun2015}).

\section{Results}

\begin{figure}[!ht]
\centering
\subfloat[]{\includegraphics[width=0.45\textwidth]{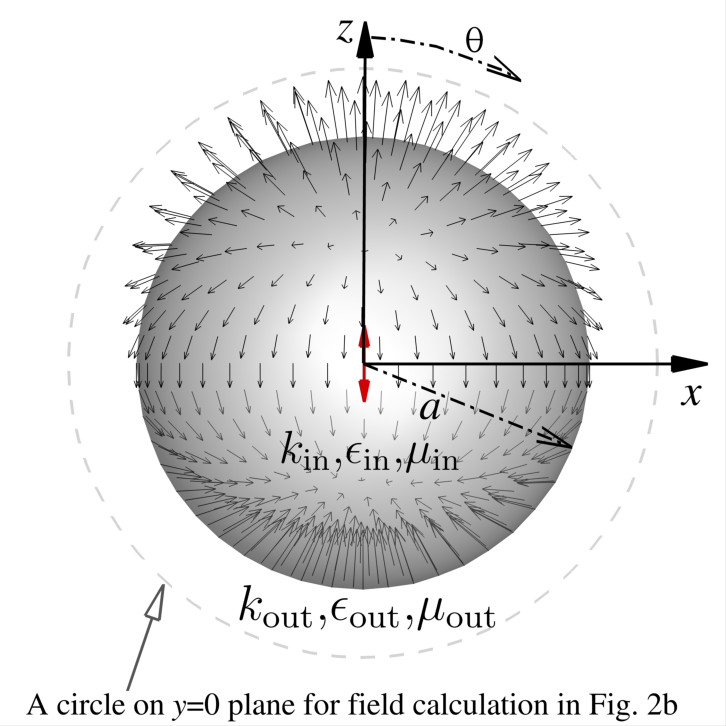}}
\subfloat[]{\includegraphics[width=0.45\textwidth]{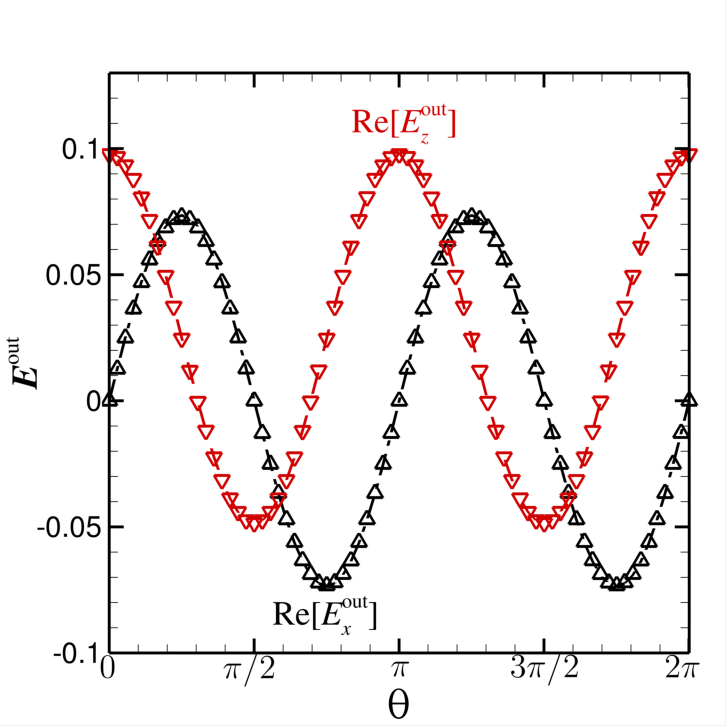}} \\
\subfloat[]{\includegraphics[width=0.5\textwidth]{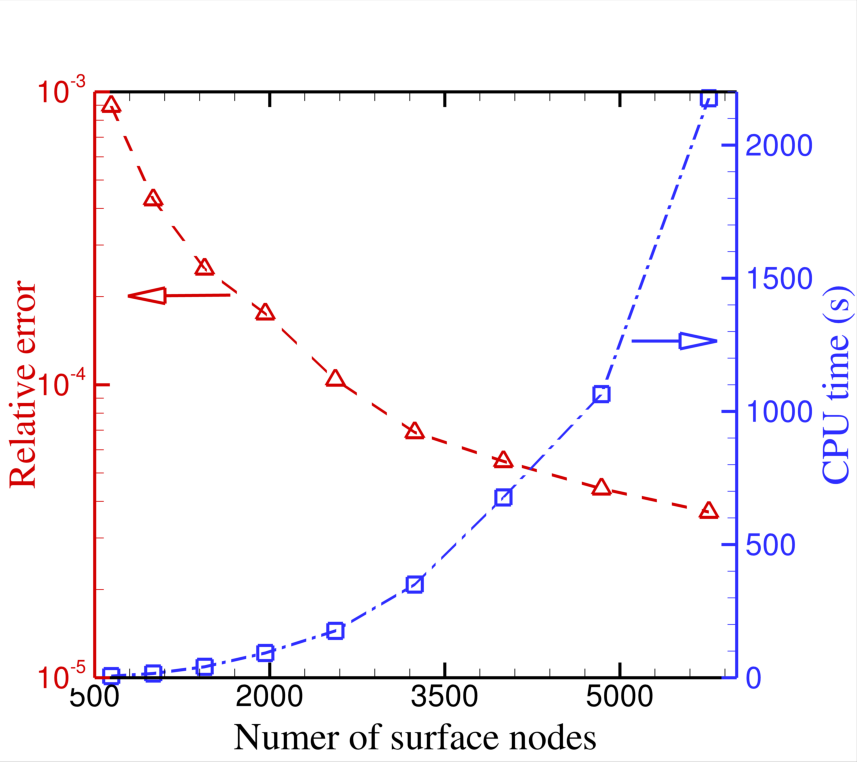}}
\caption{Comparison between the analytical results for a dielectric sphere with a dipole at its center placed in an outer infinite domain and the numerical simulation. The dipole is placed at the center of the sphere, see (a). The parameters used are $k_{\rmout}a = 1$, $k_{\rmin}a=2$, $\epsilon_{\rmin}/\epsilon_{\rmout}=4$ and $\mu_{\rmin}=\mu_{\rmout}=1$. 
Excellent agreement is obtained with the analytical result, see (b). 
In (a) the electric field vectors of the outer domain are shown as well on the surface. (c) Convergence study and computational time. Even with 642 nodes the relative error is only 0.1 percent.} \label{Fig:2anacompare}
\end{figure}

\subsection{Verification with Mie alike theory}

The numerical framework was tested with an electric dipole placed at the centre of a dielectric sphere for which an analytical solution exists. The radius is indicated with $a$, and the physical parameters are chosen to be $k_{\rmout}a = 1$, $k_{\rmin}a=2$, $\epsilon_{\rmin}/\epsilon_{\rmout}=4$ and $\mu_{\rmin}=\mu_{\rmout}=1$. The number of nodes is 642 and 320 quadratic triangular elements were used to represent the sphere surface in the simulation with the dipole moment being $\bs{p}= (0,0,1)$. The analytical solution is described in Appendix~\ref{App:AnalyticSolution}. We checked the numerical solutions of $\bs E$ on both sides of the surface and they show excellent agreement with the theory (not shown here). We also chose a circle with radius $1.2a$ at the plane $y=0$ to confirm that the electric field is correct in the domain close to the sphere. 72 sample locations were seeded along that circle with equal angular interval of polar angle $\theta$ where sample 1 corresponds to $\theta = 0$. The results are shown in Fig.~\ref{Fig:2anacompare}b, where the numerical calculations of the real parts of $E_x^{\rmout}$ (black symbols) and $E_z^{\rmout}$ (red symbols) are plotted as a function of the polar angle $\theta$ (as indicated in Fig.~\ref{Fig:2anacompare}a) and compared to their theoretical counterparts (lines). The theoretical and numerical solutions are virtually identical.
In Fig.~\ref{Fig:2anacompare}c, we investigated the convergence performance of our method with respect to the example demonstrated in Fig.~\ref{Fig:2anacompare}b. As we can see, as the number of surface nodes increases, the average relative error between the analytical solution decreases and the computational time\footnote{All the cases in Fig.~\ref{Fig:2anacompare}c were performed on a MacBook Pro with 2.6 GHz 6-Core Intel Core i7 processor and 32 GB 2400 MHz DDR4 Memory without parallel computation.} increases. The relative error in that figure is defined as 
\begin{align}
    \text{Relative error} = \frac{\sum_{1}^{72} |\bs{E}^{\rmout}_{\text{ana}}-\bs{E}^{\rmout}_{\text{num}}|}{\sum_{1}^{72} |\bs{E}^{\rmout}_{\text{ana}}|}
\end{align}
where the subscript `ana' stands for the analytical solution and `num' for the numerical solution.

Some other tests to examine the correctness and accuracy of the constructed non-singular field only boundary element framework were also performed: if the surface $S$ in Eq.~(\ref{eq:nsbim}) is very small or if this surface is situated far away from both the dipole and the observation point, then the electric field goes back to the field of the undisturbed dipole as it should, because both integrals in Eq.~(\ref{eq:nsbim}) will disappear for these cases. 

These tests give us the confidence that the constructed framework is correct and robust. In the next sections we will investigate some more physically interesting examples. 

\subsection {Nano optical-antenna}\label{sec:nanoantenna}

\begin{figure}[t]
\centering
\subfloat[$\bs{p}=\left(\frac{\sqrt{2}}{2},0,-\frac{\sqrt{2}}{2}\rmi\right)$]{\includegraphics[width=0.25\textwidth]{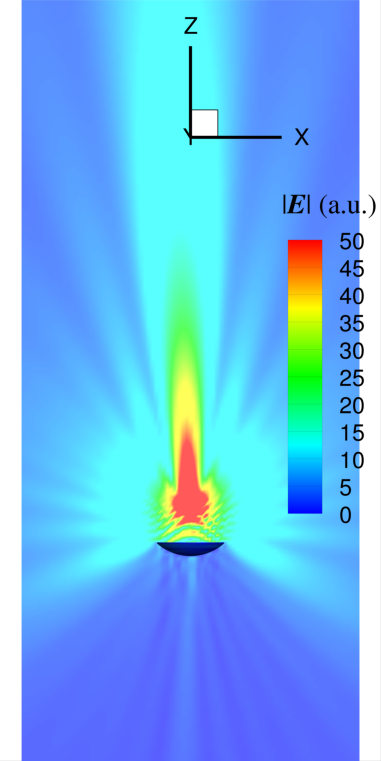}}
\subfloat[$\bs{p}=(1,0,0)$]{\includegraphics[width=0.25\textwidth]{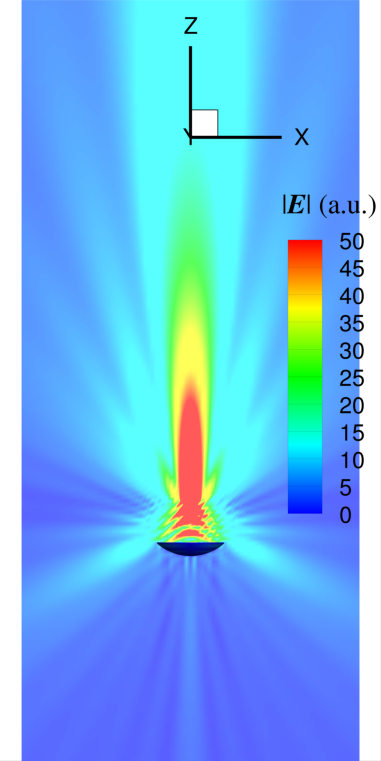}}
\subfloat[$\bs{p}=\left(\frac{\sqrt{2}}{2},0,\frac{\sqrt{2}}{2}\rmi\right)$]{\includegraphics[width=0.25\textwidth]{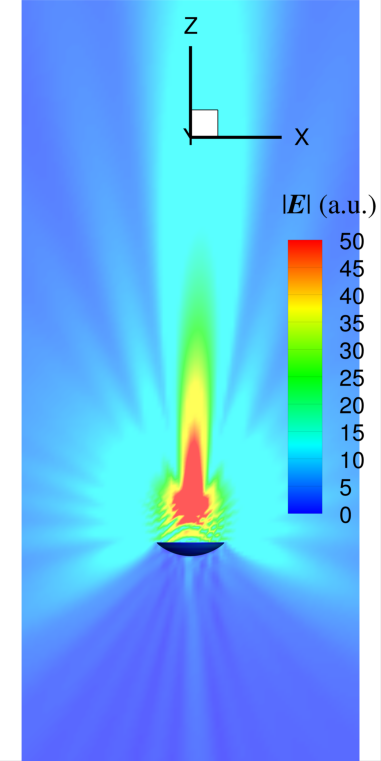}}
\subfloat[$\bs{p}=(0,0,\rmi)$]{\includegraphics[width=0.25\textwidth]{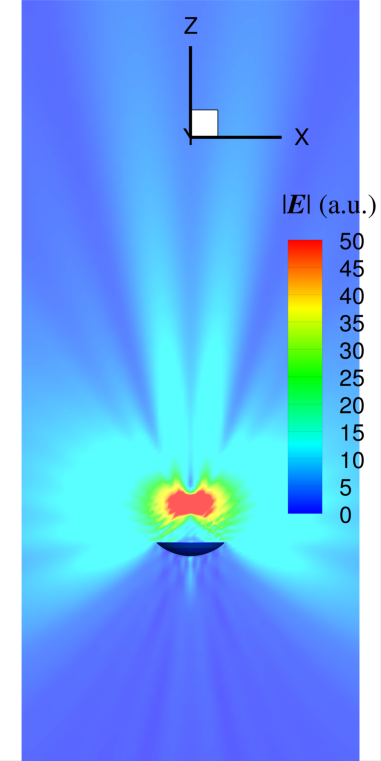}}
\caption{A dynamic electromagnetic dipole placed at the focal point of a parabolic mirror (PEC) with $ka=20$. In (a) and (c) the dipole is circular polarized with $\boldsymbol p = (1,0,-\rmi)\sqrt 2 /2$ in (a) and oppositely circular polarized with $\boldsymbol p = (1,0,\rmi)\sqrt 2 /2$ in (c). Note that the direction of the `beam' is slightly changed to the left in (a) and to the right in (c). In (b) and (d) the dipole is linearly polarized in the $x$-direction as $\boldsymbol p = (1,0,0)$ and along the $z$-direction with $\boldsymbol p = (0,0,\rmi)$, respectively. When the dipole is polarized along the axis of the mirror ($z$-direction in d), virtually no focused energy transfer is observed. } \label{Fig:3bowl}
\end{figure}
An optical antenna can be designed, synthesised and used to efficiently convert the freely propagating optical radiation to localised energy to enhance the interaction between light and mesoscopic structures~\cite{Bharadwaj2009}. In analogy to the radio antenna design, we investigate the interaction between a dynamic electric dipole and a parabolic nano-antenna transmitter device, as shown in Fig.~\ref{Fig:3bowl}. The shape and location of the nano parabolic transmitter are given by $(x,y,z) = (a \sin{\theta}\cos{\varphi}, \, a \sin{\theta}\sin{\varphi}, \, 0.1a (\cos{\theta}-1)+0.3\sin^2{\theta})$ with $\theta$ the polar angle measured from the $z$-axis, $\varphi$ the azimuthal angle and $a$ the characteristic size of the nano-antenna. The dynamic dipole is positioned at the focal point of this antenna at $(0, \, 0, \, 1.35a)$ which is along the symmetry axis of the parabolic nano-antenna ($z$-axis). In this example, we chose the characteristic non-dimensional value $ka=20$. For such a $ka$, when the light wavelength is $\lambda = 550 \, \text{nm}$, the corresponding characteristic size of the parabolic nano-antenna would be $a=1.75 \, \mu \text{m}$. Also, we assume that this antenna is well coated to be a perfect mirror (PEC) when the surrounding medium is air ($n_{\text{air}} = 1$), and use the theory and numerical method demonstrated in Sec.~\ref{sec:pec} to perform the calculations. For this test, 4002 nodes and 2000 quadratic triangular surface elements were used on the antenna surface.

In Fig.~\ref{Fig:3bowl}(b) and (d), the distribution of the electric field magnitude in the area surrounding the nano-antenna when a linearly polarised dipole interacts with it are shown. When the direction of dipole moment, $\bs{p}$, is along the symmetry axis of the antenna, the conversion of the energy of the free radiation of the dipole is very low as shown in Fig.~\ref{Fig:3bowl}(d). On the other hand, when $\bs{p}$ is perpendicular to the symmetry axis of the parabolic nano-antenna, a strong localised energy flow is formed as shown in Fig.~\ref{Fig:3bowl}(b). For the circularly polarised dipole examples of Fig.~\ref{Fig:3bowl}(a) and (c), other than the obvious conversion of energy, note that an interesting phenomenon is the changing direction of the energy flow depending on the direction of the circular polarisation. Such a change of the direction can illustrate information of the spin angular momentum of the dipole. 

\subsection{Near field gold (Au) probes}\label{sec:nanoprobe}

\begin{figure}[t]
\centering
\subfloat[$\bs{p}=(1,0,0)$]{\includegraphics[width=0.28\textwidth]{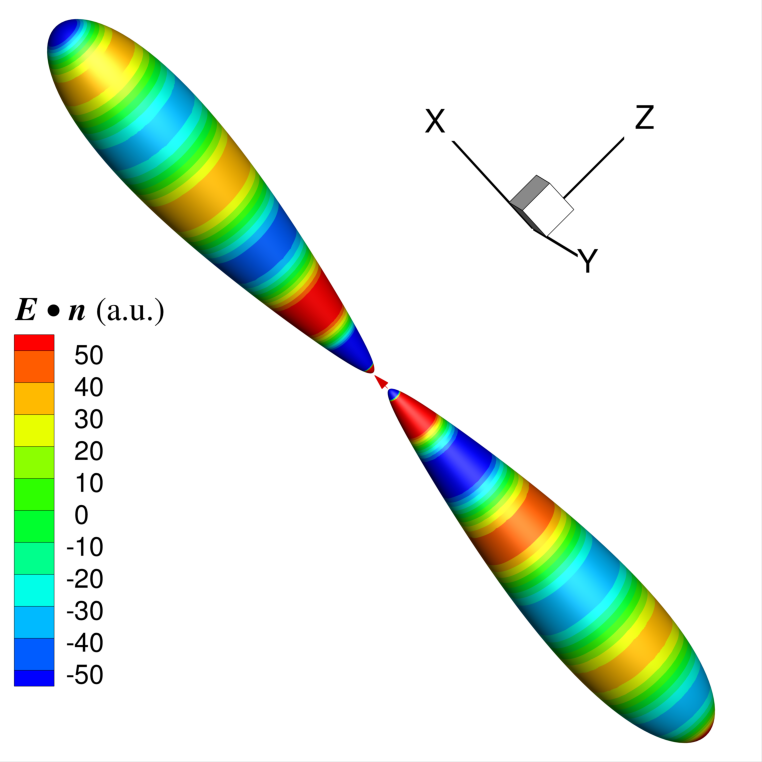}} \quad
\subfloat[$\bs{p}=(1,0,0)$]{\includegraphics[width=0.28\textwidth]{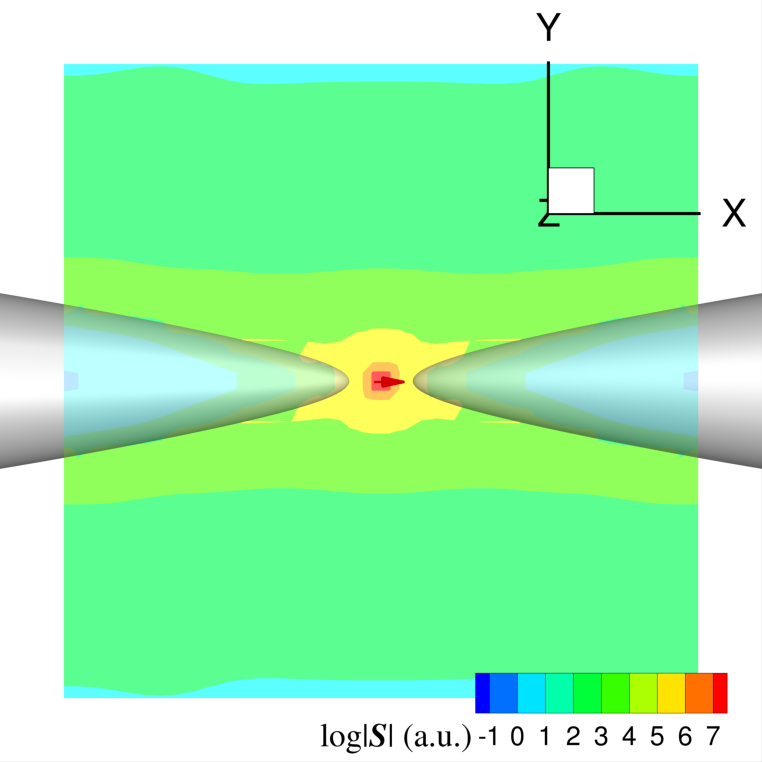}} \quad
\subfloat[$\bs{p}=(1,0,0)$]{\includegraphics[width=0.28\textwidth]{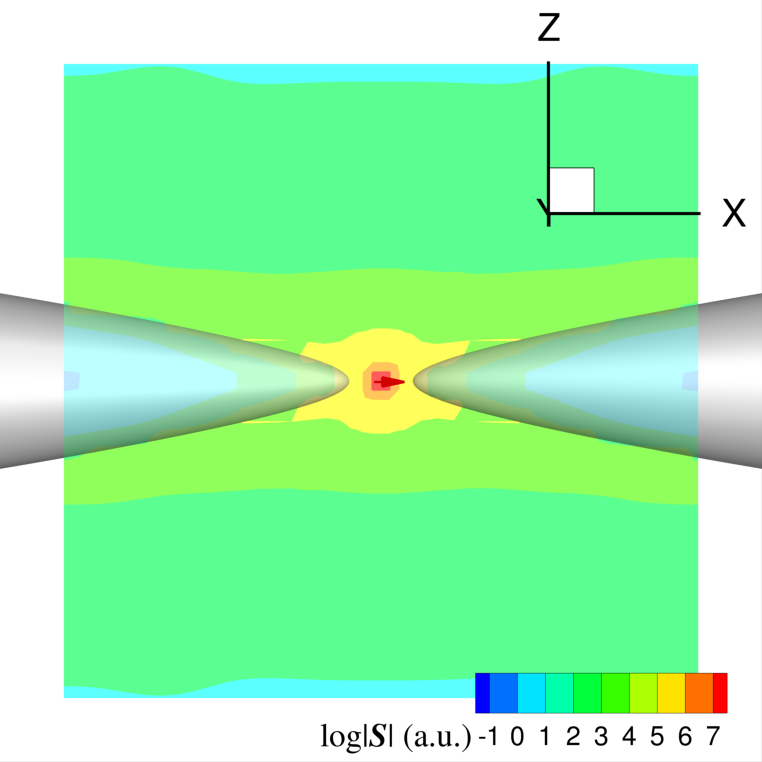}} \\
\subfloat[$\bs{p}=(0,0,1)$]{\includegraphics[width=0.28\textwidth]{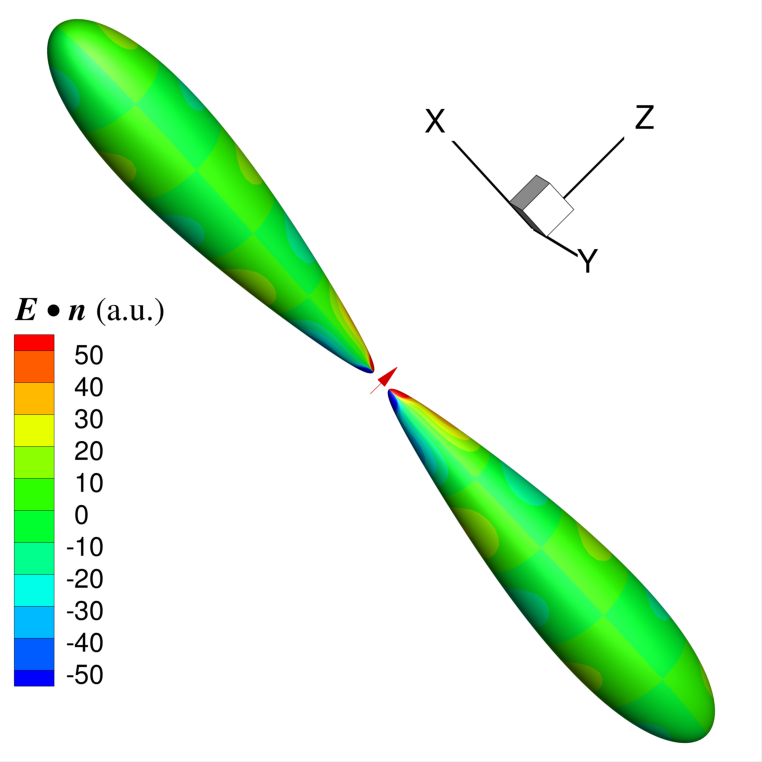}} \quad
\subfloat[$\bs{p}=(0,0,1)$]{\includegraphics[width=0.28\textwidth]{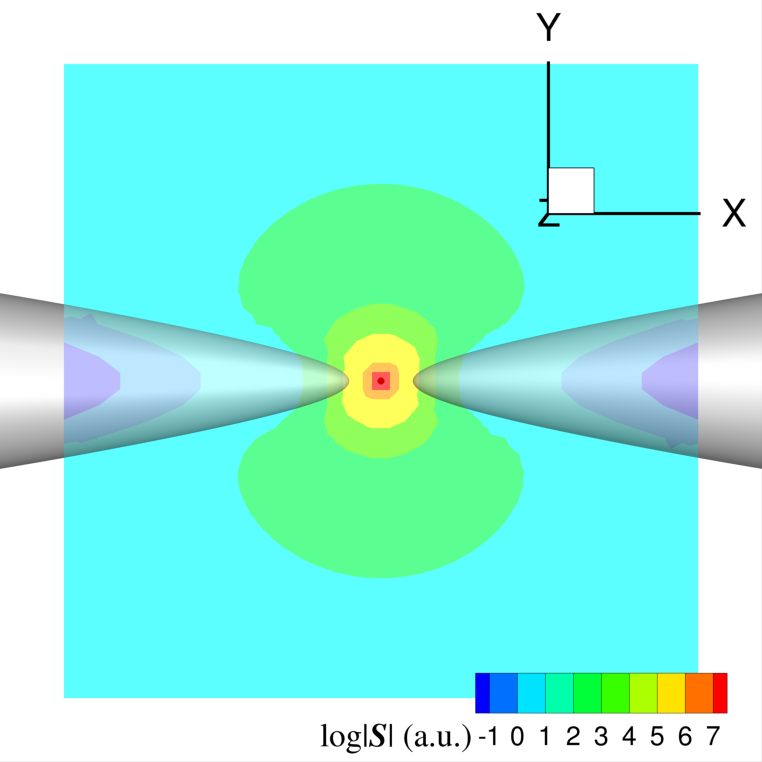}} \quad
\subfloat[$\bs{p}=(0,0,1)$]{\includegraphics[width=0.28\textwidth]{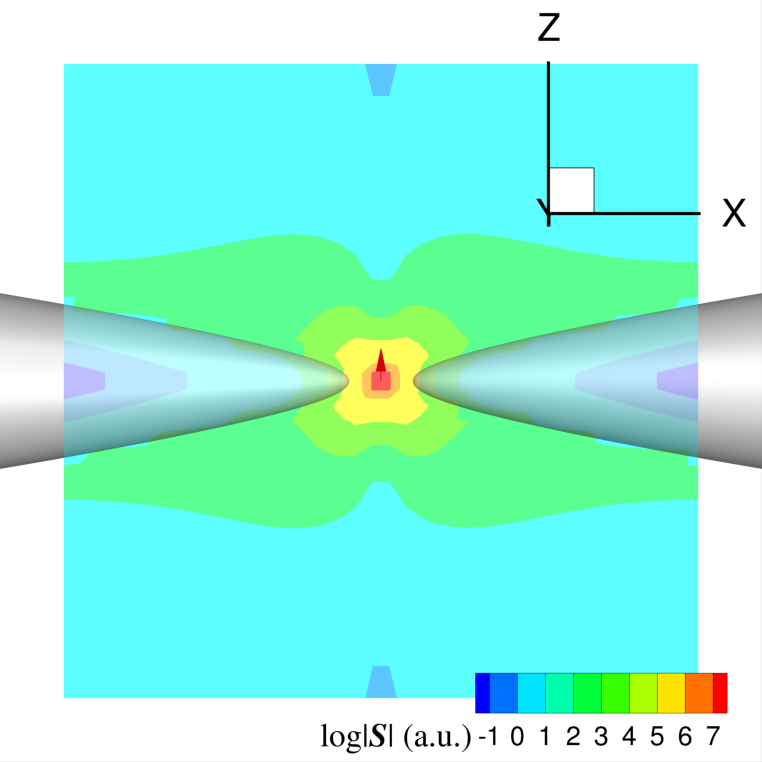}}
\caption{Two Au probes in air, with an electric dipole located exactly in the center between them ((a) and (d)). The parameters are $n_{\text{Au}} = 0.19 + \rmi 4.71$, $n_{\text{air}}=1.00$ and $\lambda = 800 \, \text{nm}$. For this case, 5762 nodes connected by 2880 quadratic triangular surface elements have been used on the surface of each probe. In the zoomed-in areas of (b) and (e) the Poynting vector $\boldsymbol S$ is plotted on the plane given by $z=0$ and in (b) and (d) on the plane $y=0$.} \label{Fig:4Auprobes}
\end{figure}

Duan et al.\cite{Duan2012} showed that classical electromagnetic calculations are capable of capturing the essential physics of a nanoplasmonic system even down to nanometer scale sized gaps and bridges. Near field optical probes, such as metal tips, are one of the key components for near field microscopes in nano-optics~\cite{Novotny2006}. The sensitivity and resolution of such a device is strongly dependent on the field distribution around the tip of the probe. We simulate the interaction between a dynamic electric dipole with two Au probes. The dipole is located at the origin of the reference coordinate system which radiates light with wavelength $\lambda = 800 \, \text{nm}$. The shape of the Au probe is described by: $x = a \sin{\theta}\cos{\varphi} [1+2\cos^2{(\theta/2)}]/10$, $y = a \sin{\theta}\sin{\varphi} [1+2\cos^2{(\theta/2)}]/10$ and $z = a\cos{\theta}$ where $\theta$ is the polar angle measured from the $z$-axis, $\varphi$ is the azimuthal angle and $a$ is the characteristic size of the probe. As a result, the tip of the Au probe points up along the $z$-axis. In this particular example, we use two such Au probes with length of 2 $\mu$m ($a=1\, \mu\text{m}$). We rotate one probe $90^{\text{o}}$ around the $y$-axis clockwise and the other counter-clockwise to make their symmetry axis along the $x$-axis with their tips pointing to each other. We then position the probes along the $x$-axis with the separation between the tips as $100 \, \text{nm}$ which implies the distance between the dipole to the tip of one probe to be $50 \, \text{nm}$. With such a gap size, classical electromagnetic phenomena are dominant, whereas quantum effects~\cite{Tserkezis2017} are deemed not significant.

We consider two scenarios: when the electric dipole moment is along the axis of the tips, $\bs{p} = (1,0,0)$ and when it is perpendicular to the axis of the tips, $\bs{p} = (0,0,1)$. In both scenarios, the surrounding medium is set to be air with $n_{\text{air}} = 1.00$. When $\lambda = 800 \, \text{nm}$, the complex refractive index of Au is $n_{\text{Au}} = 0.19 + \rmi 4.71$. As such, the coupling effects between the internal and external electromagnetic fields of the Au probes were solved using the framework described in Sec.~\ref{sec:dielectrics}. 

In Figs.~\ref{Fig:4Auprobes}a and Fig.~\ref{Fig:4Auprobes}d, the distribution of the instantaneous induced surface charge (which is equivalent to the normal component of the surface electric field: $Re(\boldsymbol E \cdot \boldsymbol n)$) is presented. When the electric dipole moment is along the axis of the tips, the electric polarisation of the radiation light from that dipole is parallel to the tip axis which induces axis-symmetric surface charge density with the highest amplitude at the ends of the tips, as demonstrated in Fig.~\ref{Fig:4Auprobes}a. However, when the electric dipole moment is perpendicular to the axis of the tips, the electric polarisation of the radiation light from that dipole is also perpendicular to the tip axis. This leads to opposite surface charges on the diametrically opposed surface points on the probes with neutral foremost tips, as demonstrated in Fig.~\ref{Fig:4Auprobes}d. The difference is also reflected by the energy propagation within the probes. We compared the amplitude of the Poynting vector $\bs{S} = 1/2 \text{Re}[\bs{E}\times (\boldsymbol{H})^{*}]$ (with $^*$ indicating the conjugate) for those two scenarios in Figs.~\ref{Fig:4Auprobes}b,c and Figs.~\ref{Fig:4Auprobes}e,f, respectively, where $\bs{H}$ was solved by the framework illustrated in Sec.~\ref{sec:discussion}. We can see that the electromagnetic energy penetrates deeper into the Au probes when the electric dipole moment is along the tip axis than when it is perpendicular to that axis.

\section{Discussion}\label{sec:discussion}

The focus of the previous sections was on the electric field. Given the symmetry between the $\bs{E}$ and $\bs{H}$ field in the Maxwell equations, in the domain that excluding the dipole the magnetic field $\bs H$ also satisfies the vector Helmholtz equation and the divergence free condition:
\begin{subequations}\label{eq:MaxwellEq}
    \begin{align}
        &\nabla^2 \bs{H} + k^2 \bs{H} = \bs{0}, \label{eq:HelmH}\\
        &\nabla \cdot \bs{H} = 0. \label{eq:divH}
    \end{align}
\end{subequations}
As such, we can equally well use the same framework from Sec.~\ref{sec:dielectrics} for the magnetic field for the interaction between an electric dipole and a dielectric object when Eq.~(\ref{eq:H_Edipole}) is used. 

Though our method is demonstrated with the interaction between a single electric dipole in the previous section, it is obvious that such a method can be extended to multiple dipoles straightforwardly. Moreover, the idea to calculate $\bs{I}^{\rmEd}$ in Appendix~\ref{app:integral} is a general one that can be used for other point sources, such as magnetic dynamic dipoles and dynamic quadrupoles. Since higher order multipoles can alternatively be represented by a few dipoles very close to each other, we concentrate on dipoles in this work. Besides electric dipoles, there also exist magnetic dipoles. For a magnetic dipole, the electric and magnetic fields are, respectively~\cite{jackson1999classical} 
\begin{subequations}\label{eq:EH_Hdipole}
\begin{align}
    \bs{E}^{\rmHd} =  \frac{\rmi \omega \mu_0 \mu_{r}}{4\pi } \left \{ \nabla \left[\frac{\exp( \rmi k |\bs{x} - \bs{x}_{d}|)}{|\bs{x} - \bs{x}_{d}|}\right] \times \bs{m} \right\}, \label{eq:E_Hdipole}\\
    \bs{H}^{\rmHd} = \frac{1}{4\pi } \nabla \times \left \{ \nabla \left[\frac{\exp( \rmi k |\bs{x} - \bs{x}_{d}|)}{|\bs{x} - \bs{x}_{d}|}\right] \times \bs{m} \right\} \label{eq:H_Hdipole}
\end{align}
\end{subequations}
where $\bs{m}$ 
is the magnetic dipole moment. 

In summary, the electric and magnetic fields on dielectric scatterer surfaces driven by the multiple dynamic electric and magnetic dipoles in the unbounded external domain are
\begin{subequations}\label{eq:nsbimE}
    \begin{align}
        &4\pi \bs{E}^{\rmout}(\boldsymbol x_0) + \int_{S}  \left[\bs{E}^{\rmout}(\boldsymbol x)  H^{\rmout}_k(\bs{x},\bs{x}_0) -\bs{F}^{\rmout}_{E}(\boldsymbol x)  H_0(\bs{x},\bs{x}_0)\right] \rmd S(\bs{x}) \nonumber \\
        = & \int_{S} \left[\frac{\del \bs{E}^{\rmout}(\boldsymbol x) } {\del n}  G^{\rmout}_k(\bs{x},\bs{x}_0) - \frac{\del \bs{F}^{\rmout}_{E}(\boldsymbol x) } {\del n}  G_0(\bs{x},\bs{x}_0) \right] \rmd S(\bs{x}) \nonumber \\ & + \sum_{i = 1}^{M^{\rmout}_{E}} \frac{1}{\epsilon_0 \epsilon_{\rmout} } \nabla \times \left \{ \nabla  G^{\rmout}_{k}(\bs{x}_0,\bs{x}_{d,i}) \times \bs{p}_{i} \right\} + \sum_{j = 1}^{M^{\rmout}_{H}} (\rmi \omega \mu_0 \mu_{\rmout}) \left \{ \nabla G^{\rmout}_{k}(\bs{x}_0,\bs{x}_{d,j}) \times \bs{m}_{j} \right\}, \label{eq:nsbimE_int} \\
        & \nabla \cdot \bs{E}^{\rmout} = 0 \qquad \text{on scatterer surfaces};
    \end{align}
\end{subequations} 
\begin{subequations}\label{eq:nsbimH}
    \begin{align}
        &4\pi \bs{H}^{\rmout}(\boldsymbol x_0) + \int_{S}  \left[\bs{H}^{\rmout}(\boldsymbol x)  H^{\rmout}_k(\bs{x},\bs{x}_0) -\bs{F}^{\rmout}_{H}(\boldsymbol x)  H_0(\bs{x},\bs{x}_0)\right] \rmd S(\bs{x}) \nonumber \\
        = & \int_{S} \left[\frac{\del \bs{H}^{\rmout}(\boldsymbol x) } {\del n}  G^{\rmout}_k(\bs{x},\bs{x}_0) - \frac{\del \bs{F}^{\rmout}_{H}(\boldsymbol x) } {\del n}  G_0(\bs{x},\bs{x}_0) \right] \rmd S(\bs{x}) \nonumber \\ & + \sum_{i = 1}^{M^{\rmout}_{E}} \frac{\omega}{\rmi} \left \{ \nabla  G^{\rmout}_{k}(\bs{x}_0,\bs{x}_{d,i}) \times \bs{p}_{i} \right\} + \sum_{j = 1}^{M^{\rmout}_{H}} \nabla \times \left \{ \nabla G^{\rmout}_{k}(\bs{x}_0,\bs{x}_{d,j}) \times \bs{m}_{j} \right\}, \label{eq:nsbimH_int} \\
        & \nabla \cdot \bs{H}^{\rmout} = 0 \qquad \text{on scatterer surfaces}.
    \end{align}
\end{subequations}
where $M^{\rmin}_{E}$ and $M^{\rmin}_{H}$ are, respectively, the number of dynamic electric and magnetic dipoles in the external domain, and $\bs{F}^{\rmout}_{E}(\bs{x})$ $\bs{F}^{\rmout}_{H}(\bs{x})$ are, respectively, 
\begin{subequations}\label{eq:nsbim_psiEH}
   \begin{align}
   \bs{F}^{\rmout}_{E}(\boldsymbol x) &= \bs{E}^{\rmout}(\boldsymbol x_0) + [\boldsymbol n(\boldsymbol x_0) \cdot (\boldsymbol x - \boldsymbol x_0)] \frac{\del \bs{E}^{\rmout}(\boldsymbol x_0)}{\del n} \\
    \bs{F}^{\rmout}_{H}(\boldsymbol x) &= \bs{H}^{\rmout}(\boldsymbol x_0) + [\boldsymbol n(\boldsymbol x_0) \cdot (\boldsymbol x - \boldsymbol x_0)] \frac{\del \bs{H}^{\rmout}(\boldsymbol x_0)}{\del n}.
\end{align} 
\end{subequations}

For the internal field on the dielectric scatterer surface, we have
\begin{subequations}\label{eq:nsbimEin}
    \begin{align}
        &\int_{S}  \left[\bs{E}^{\rmin}(\boldsymbol x)  H^{\rmin}_k(\bs{x},\bs{x}_0) -\bs{F}^{\rmin}_{E}(\boldsymbol x)  H_0(\bs{x},\bs{x}_0)\right] \rmd S(\bs{x}) \nonumber \\
        = & \int_{S} \left[\frac{\del \bs{E}^{\rmin}(\boldsymbol x) } {\del n}  G^{\rmin}_k(\bs{x},\bs{x}_0) - \frac{\del \bs{F}^{\rmin}_{E}(\boldsymbol x) } {\del n}  G_0(\bs{x},\bs{x}_0) \right] \rmd S(\bs{x}) \nonumber \\ & - \sum_{i = 1}^{M^{\rmin}_{E}} \frac{1}{\epsilon_0 \epsilon_{\rmin} } \nabla \times \left \{ \nabla  G^{\rmin}_{k}(\bs{x}_0,\bs{x}_{d,i}) \times \bs{p}_{i} \right\} - \sum_{j = 1}^{M^{\rmin}_{H}} (\rmi \omega \mu_0 \mu_{\rmin}) \left \{ \nabla G^{\rmin}_{k}(\bs{x}_0,\bs{x}_{d,j}) \times \bs{m}_{j} \right\}, \label{eq:nsbimEin_int} \\
        & \nabla \cdot \bs{E}^{\rmin} = 0 \qquad \text{on scatterer surfaces};
    \end{align}
\end{subequations} 
\begin{subequations}\label{eq:nsbimHin}
    \begin{align}
        &\int_{S}  \left[\bs{H}^{\rmin}(\boldsymbol x)  H^{\rmin}_k(\bs{x},\bs{x}_0) -\bs{F}^{\rmin}_{H}(\boldsymbol x)  H_0(\bs{x},\bs{x}_0)\right] \rmd S(\bs{x}) \nonumber \\
        = & \int_{S} \left[\frac{\del \bs{H}^{\rmin}(\boldsymbol x) } {\del n}  G^{\rmin}_k(\bs{x},\bs{x}_0) - \frac{\del \bs{F}^{\rmin}_{H}(\boldsymbol x) } {\del n}  G_0(\bs{x},\bs{x}_0) \right] \rmd S(\bs{x}) \nonumber \\ & - \sum_{i = 1}^{M^{\rmin}_{E}} \frac{\omega}{\rmi} \left \{ \nabla  G^{\rmin}_{k}(\bs{x}_0,\bs{x}_{d,i}) \times \bs{p}_{i} \right\} - \sum_{j = 1}^{M^{\rmin}_{H}} \nabla \times \left \{ \nabla G^{\rmin}_{k}(\bs{x}_0,\bs{x}_{d,j}) \times \bs{m}_{j} \right\}, \label{eq:nsbimHin_int} \\
        & \nabla \cdot \bs{H}^{\rmin} = 0 \qquad \text{on scatterer surfaces}.
    \end{align}
\end{subequations}
where $M^{\rmin}_{E}$ and $M^{\rmin}_{H}$ are, respectively, the number of dynamic electric and magnetic dipoles inside the scatterer, and $\bs{F}^{\rmin}_{E}(\bs{x})$ $\bs{F}^{\rmin}_{H}(\bs{x})$ are, respectively, 
\begin{subequations}\label{eq:nsbim_psiEHin}
   \begin{align}
   \bs{F}^{\rmin}_{E}(\boldsymbol x) &= \bs{E}^{\rmin}(\boldsymbol x_0) + [\boldsymbol n(\boldsymbol x_0) \cdot (\boldsymbol x - \boldsymbol x_0)] \frac{\del \bs{E}^{\rmin}(\boldsymbol x_0)}{\del n} \\
    \bs{F}^{\rmin}_{H}(\boldsymbol x) &= \bs{H}^{\rmin}(\boldsymbol x_0) + [\boldsymbol n(\boldsymbol x_0) \cdot (\boldsymbol x - \boldsymbol x_0)] \frac{\del \bs{H}^{\rmin}(\boldsymbol x_0)}{\del n}.
\end{align} 
\end{subequations}

It is worth emphasizing that the sums with $M^{\rmout}_{E}$, $M^{\rmout}_{H}$, $M^{\rmin}_{E}$ and $M^{\rmin}_{H}$ will end up on the right hand side of the resulting numerical matrix system, and will thus require barely any noticeable additional computational resources, regardless how many dipoles are situated in the domain. 

It is also worth pointing out that, for the examples shown in Secs.~\ref{sec:nanoantenna} and \ref{sec:nanoprobe}, we used the surface meshes that were transformed from the mesh of a sphere, which is just for our own convenience. Of course, for practical applications, our method can accept any surface mesh of a closed object.

\section{Conclusions}
In this work, we illustrate a non-singular field-only surface integral method for the interactions between electric and magnetic dipoles with mesoscale structures. We have shown that it is relatively easy to implement such volume sources, electric or magnetic dipoles, in our framework to simulate electromagnetic phenomena. The results were verified with a Mie alike theoretical solution. Two examples were given; one with a parabolic nano-antenna with a dipole at its focal point and one with two gold (Au) probes situated around a dipole. Our framework can also deal with multiple dipoles straightforwardly without any substantial increase in the computational requirements. We believe that our method can make a good contribution to addressing one of the key problems in nano-optics to determine the electromagnetic field distributions near and on mesoscale structures. Also, considering that the Maxwell equations describe electromagnetic phenomena for length scales ranging over more than 10 orders of magnitude, our framework can also be used beyond nano-optical topics.








\medskip
\textbf{Acknowledgements} \par 
Q.S. acknowledges the support from the Australian Research Council grants DE150100169, FT160100357 and CE140100003. This research was partially undertaken with the assistance of resources from the National Computational Infrastructure (NCI Australia), an NCRIS enabled capability supported by the Australian Government (Grant No. LE160100051).

\appendix

\section{Derivation of the surface integral equations} \label{app:integral}

In this appendix the derivations of the integral equations of Eq.~(\ref{eq:nsbim}) are given, including the desingularization and the treatment of the singular dipole terms. The standard boundary integral equation for each component of the electric field $E_x$, $E_y$ and $E_z$ is
\begin{equation} \label{eq:bimA1}
    \begin{aligned}
C(\bs{x}_0) \bs{E}(\boldsymbol x_0) + \int_{S}  \left[\bs{E}(\boldsymbol x)  H_k(\bs{x},\bs{x}_0) 
- \frac{\del \bs{E}(\boldsymbol x) } {\del n}  G_k(\bs{x},\bs{x}_0)  \right] \rmd S(\bs{x}) + \bs{I}^{\rmEd} =\bs 0,
\end{aligned}
\end{equation}
where we have excluded the integrals over a small sphere containing the dipole with surface $S_d$ (see Fig.~\ref{Fig:1sketch}) as 
\begin{equation} \label{eq:bimA2}
    \begin{aligned}
\bs{I}^{\rmEd} = \int_{S_d}  \left[\bs{E}(\boldsymbol x)  H_k(\bs{x},\bs{x}_0)   
- \frac{\del \bs{E}(\boldsymbol x) } {\del n}  G_k(\bs{x},\bs{x}_0)  \right] \rmd S(\bs{x}).
\end{aligned}
\end{equation}
In Eq.~(\ref{eq:bimA1}), $C(\bs{x}_0)$ is the solid angle at $\bs{x}_0$. Numerically, due care needs to be taken in calculating the integrals since the integrals become singular at $\bs x = \bs x_0$ because of the presence of $H_k$ and $G_k$ in Eq.~(\ref{eq:bimA1}). However, these singularities can be analytically removed by subtracting other integrals with the same singular behavior. One possible choice would be:
\begin{equation} \label{eq:bimA3}
    \begin{aligned}
C(\bs{x}_0) \bs{F}(\boldsymbol x_0) + \int_{S}  \left[\bs{F}(\boldsymbol x)  H_0(\bs{x},\bs{x}_0) 
- \frac{\del \bs{F}(\boldsymbol x) } {\del n}  G_0(\bs{x},\bs{x}_0)  \right] \rmd S(\bs{x}) =\bs 0,
\end{aligned}
\end{equation}
where $\bs{F}(\bs{x})$ satisfies the Laplace equation as $\nabla^2 \bs{F}(\bs{x}) = \bs{0}$, and $G_0$ and $H_0$ are the Green's function for the Laplace equation and its normal derivative. $G_0$ has the same $1/r$ singular behavior as $G_k$. If we choose $\bs F(\bs x)= \bs E(\bs x_0) + \bs n(\bs x_0) \cdot (\bs x - \bs x_0) \partial \bs E (\bs x_0)/\partial n$, such that $\bs F$ approaches $\bs E(\bs x_0)$ when $\bs x$ approaches $\bs x_0$ and the normal derivative $\lim_{\bs x \rightarrow \bs x_0}\partial F (\bs x_0)/\partial n = \lim_{\bs x \rightarrow \bs x_0}\bs n(\bs x)\cdot \bs n( \bs x_0) \partial E (\bs x_0)/\partial n = \partial E (\bs x_0)/\partial n$. By subtracting Eq.~(\ref{eq:bimA3}) from Eq.~(\ref{eq:bimA1}), we end up with Eq.~(\ref{eq:nsbim}) of the main text. It is worth noting that the term with the solid angle $C(\bs{x}_0)$ disappears. For an unbounded external domain, however, a term $4\pi \bs E (\bs x_0)$ will appear due to the first term of the $\bs{F}(\bs{x})$ function (the integral over a domain at infinity will no longer be zero but will generate this  $4\pi \bs E (\bs x_0)$ term. On the other hand, for internal domains, this term will not appear, see Eq.~(\ref{eq:nsbimIN2}) for example. 

The last term on the right hand side of Eq.~(\ref{eq:nsbim}), $\bs{I}^{\rmEd}$, which is defined in Eq.~(\ref{eq:nsbim_Id}), can be calculated analytically as follows.
Since the size of $S_{d}$ is small, the electric field on $S_{d}$ is dominated by the electric dipole. As such, when Eq.~(\ref{eq:E_Edipole}) is used, $\bs{I}^{\rmEd}$ can be calculated as
\begin{equation}
\begin{aligned}
    \bs{I}^{\rmEd} =& \quad \; \frac{1}{4\pi \epsilon_0 \epsilon_r } \lim_{r_d \rightarrow 0}\int_{S_{d}} H_{k}(\bs{x},\bs{x}_0) \, \nabla \times \left \{ \nabla \left[\frac{\exp( \rmi k |\bs{x} - \bs{x}_{d}|)}{|\bs{x} - \bs{x}_{d}|} \right] \times \bs{p} \right\} \rmd S(\bs{x})\\
      &- \frac{1}{4\pi \epsilon_0 \epsilon_r } \lim_{r_d \rightarrow 0}\int_{S_{d}}  G_{k}(\bs{x},\bs{x}_0) \, \frac{\del}{ \del n} \Bigg( \nabla \times \left \{ \nabla \left[\frac{\exp( \rmi k |\bs{x} - \bs{x}_{d}|)}{|\bs{x} - \bs{x}_{d}|}\right] \times \bs{p} \right\} \Bigg)\rmd S(\bs{x}). 
\end{aligned}
\end{equation}
Since $\nabla F(\bs x-\bs x_d)= -\nabla_{d}F(\bs x-\bs x_d)$ and $\nabla \times F(\bs x-\bs x_d)= - \nabla_{d} \times F(\bs x-\bs x_d)$ where $\nabla_d$ is the gradient with respect to $\bs{x_d}$ and $F$ any regular function, we can take $\nabla_d$ out of the integral and write :
\begin{equation}
\begin{aligned}
    \bs{I}^{\rmEd} = & \quad \; \frac{1}{4\pi \epsilon_0 \epsilon_r }   \, \nabla_d \times \left \{ \nabla_d \lim_{r_d \rightarrow 0}\int_{S_{d}} H_{k}(\bs{x},\bs{x}_0)\left[\frac{\exp( \rmi k |\bs{x} - \bs{x}_{d}|)}{|\bs{x} - \bs{x}_{d}|} \right] \rmd S(\bs{x}) \times \bs{p} \right\} \\
      & -\frac{1}{4\pi \epsilon_0 \epsilon_r } \nabla_d \times \left \{ \nabla_d \lim_{r_d \rightarrow 0}\int_{S_{d}} G_{k}(\bs{x},\bs{x}_0) \,\frac{\del}{ \del n}\left[\frac{\exp( \rmi k |\bs{x} - \bs{x}_{d}|)}{|\bs{x} - \bs{x}_{d}|}\right] \rmd S(\bs{x})\times \bs{p} \right\}. 
\end{aligned}
\end{equation}
The integrals are now easier to deal with. For instance,
\begin{equation}
    \begin{aligned}
   &\lim_{r_d \rightarrow 0} \int_{S_{d}} H_{k}(\bs{x},\bs{x}_0)\left[\frac{\exp( \rmi k |\bs{x} - \bs{x}_{d}|)}{|\bs{x} - \bs{x}_{d}|} \right] \rmd S(\bs{x}) \\ \approxeq & H_{k}(\bs{x}_d,\bs{x}_0) \lim_{r_d \rightarrow 0} \int_{S_{d}}\frac{\exp( \rmi k |\bs{x} - \bs{x}_{d}|)}{|\bs{x} - \bs{x}_{d}|}  \rmd S(\bs{x}) =0\\
    \end{aligned}
\end{equation}
since $|\bs{x} - \bs{x}_{d}| \sim 1/r_d$, but $\rmd S(\bs{x}) \sim 4 \pi r_d^2$. The second integral can be calculated using $\del /\del n = - \rmd /\rmd r_d$ and  $\del \left[\exp( \rmi k |\bs{x} - \bs{x}_{d}|)/|\bs{x} - \bs{x}_{d}|\right]/\del n = -(\rmi k r_d -1)/r_d^2$, which leads to:
\begin{equation}
    \begin{aligned}
    & \lim_{r_d \rightarrow 0}\int_{S_{d}} G_{k}(\bs{x},\bs{x}_0) \,\frac{\del}{ \del n}\left[\frac{\exp( \rmi k |\bs{x} - \bs{x}_{d}|)}{|\bs{x} - \bs{x}_{d}|}\right] \rmd S(\bs{x})  \\
    \approxeq &G_{k}(\bs{x}_d,\bs{x}_0) \lim_{r_d \rightarrow 0}\int_{S_{d}} \,\frac{\del}{ \del n}\left[\frac{\exp( \rmi k |\bs{x} - \bs{x}_{d}|)}{|\bs{x} - \bs{x}_{d}|}\right] \rmd S(\bs{x}) \\= & -G_{k}(\bs{x}_d,\bs{x}_0) \lim_{r_d \rightarrow 0} \frac{\rmi kr_d-1}{r_d^2} 4 \pi r_d^2 =4 \pi G_{k}(\bs{x}_d,\bs{x}_0).
    \end{aligned}
\end{equation}
Eventually,
\begin{equation}
\begin{aligned}
    \bs{I}^{\rmEd} &=
       -\frac{1}{\epsilon_0 \epsilon_r } \nabla_d \times \left \{ \nabla_d  G_{k}(\bs{x}_d,\bs{x}_0) \times \bs{p} \right\} = -\frac{1}{\epsilon_0 \epsilon_r } \nabla \times \left \{ \nabla  G_{k}(\bs{x}_0,\bs{x}_d) \times \bs{p} \right\} \\
       &=-\frac{\exp(\rmi k r_{0d})}{\epsilon_0 \epsilon_r r_{0d}^3}\left\{(-k^2 r_{0d}^2 - 3 \rmi k r_{0d} + 3) \frac{(\bs x_{0d}\cdot \bs p)}{r_{0d}^2} \bs x_{0d}  +(k^2 r_{0d}^2 + \rmi k r_{0d}-1)\bs p  \right\}
\end{aligned}
\end{equation}
which is the same as Eq.~(\ref{eq:Iterm}) in the main text. 

\section{Analytical solution}\label{App:AnalyticSolution}

In this appendix, the analytical solution of the field driven by an electric dipole with dipole moment $\bs{p}$ located at the center of a spherical dielectric scatter is derived. As shown in Fig.~\ref{Fig:2anacompare}a, the centre of a spherical object is set at the origin of a Cartesian coordinate system. The axis of symmetry of this spherical scatterer is chosen along the $z$-axis which aligns with the direction of the electric dipole moment $\bs{p} = (0, \, 0, \, p)$. To solve the electromagnetic field of such a system, it is most convenient to employ a spherical coordinate system, $(r, \, \theta, \, \varphi)$, which origin is at the center of the spherical scatterer and the polar angle is measured from the $z$-axis. In this spherical system, $\bs{p} = p \cos{\theta} \, \bs{e}_{r} - p \sin{\theta} \, \bs{e}_{\theta}$ where $\bs{e}_{r}$ is the unit vector along the $r$-axis and $\bs{e}_{\theta}$ the unit vector along the $\theta$-axis. 

From Eq.~(\ref{eq:EH_Edipole}), the components of the electric and magnetic field in the spherical coordinate system driven by such an electric dipole inside the sphere are
\begin{subequations}\label{eq:EM_dip_sphcoor}
    \begin{align}
        E^{\rmEd}_{r}&=\frac{1}{4\pi\epsilon_0\epsilon_2} p \left(\frac{2}{r^3} - \frac{2\rmi k_2}{r^2}\right) \exp{(\rmi k_2 r)} \cos{\theta} \\
        E^{\rmEd}_{\theta}&=\frac{1}{4\pi\epsilon_0\epsilon_2} p \left(\frac{1}{r^3} - \frac{\rmi k_2}{r^2} - \frac{k_2^2}{r} \right)\exp{(\rmi k_2r)} \sin{\theta} \\
        E^{\rmEd}_{\varphi}& = 0 \\
        H^{\rmEd}_{r} & = 0 \\
        H^{\rmEd}_{\theta} & = 0 \\
        H^{\rmEd}_{\varphi} & = \frac{k_2^3}{4\pi \epsilon_0\epsilon_2 \mu_0 \mu_2 \omega}  p \left( \frac{1}{\rmi k_2 r^2} - \frac{1}{r} \right)\exp{(\rmi k_2r)} \sin{\theta}
    \end{align}
\end{subequations}
where $k_2$ is the wavenumber, $c_2$ is the speed of light, and $\epsilon_2$ is the relative permittivity of the sphere.

Based on the Mie theory~\cite{Mie1908, hulst1981light}, when the Debye potentials $u$ and $v$ are used, we can write the electric and magnetic fields as
\begin{subequations}\label{eq:em_EH_MN}
    \begin{align}
        \bs{E} &= \bs{M}_{v} - \rmi \bs{N}_{u}, \\
        \bs{H} &= \sqrt{\frac{\epsilon_0\epsilon_r}{\mu_0 \mu_r}} (-\rmi \bs{N}_v - \bs{M}_u)
    \end{align}
\end{subequations}
with
\begin{align}
    \bs{M}_{u} = \nabla \times (\bs{x} u), \qquad \bs{M}_{v} = \nabla \times (\bs{x} v);
\end{align}
where we have used
\begin{subequations}
    \begin{align}
        \nabla \times \bs{M}_{u} & = \omega (\epsilon_0\epsilon_r\mu_0 \mu_r)^{\frac{1}{2}} \bs{N}_u, \qquad
        \nabla \times \bs{M}_{v}  = \omega (\epsilon_0\epsilon_r\mu_0 \mu_r)^{\frac{1}{2}} \bs{N}_v, \\
        \nabla \times \bs{N}_{u} & = \omega (\epsilon_0\epsilon_r\mu_0 \mu_r)^{\frac{1}{2}} \bs{M}_u, \qquad
        \nabla \times \bs{N}_{v}  = \omega (\epsilon_0\epsilon_r\mu_0 \mu_r)^{\frac{1}{2}} \bs{M}_v; 
    \end{align}
\end{subequations}
and 
\begin{align}
    \nabla \times \bs{E}  = \rmi \omega \mu_0 \mu_r \bs{H}, \qquad \nabla \times \bs{H}  =-\rmi \omega \epsilon_0\epsilon_r \bs{E}.
\end{align}
The components of $\bs{M}_{u}$ and $\bs{N}_{u}$ in spherical coordinates are:
\begin{subequations}\label{eq:MNu}
    \begin{align}
        M_{u_{r}} = 0, & \qquad M_{u_{\theta}} = \frac{1}{r\sin{\theta}} \frac{\partial (ru)}{\partial \varphi},  \qquad M_{u_{\varphi}} =-\frac{1}{r} \frac{\partial(ru)}{\partial \theta}; \\
        N_{u_{r}}  = \frac{1}{k} \frac{\partial^2(ru)}{\partial r^2} + k  r u, & \qquad
        N_{u_{\theta}} = \frac{1}{k r} \frac{\partial^2(ru)}{\partial r \partial \theta}, \qquad
        N_{u_{\varphi}} = \frac{1}{k r \sin\theta} \frac{\partial^2(ru)}{\partial r \partial \varphi}.
    \end{align}
\end{subequations}
The above formulation can also be used to get the components of $\bs{M}_{v}$ and $\bs{N}_{v}$ when the potential $u$ is replaced by the potential $v$. 

Both Debye potentials, $u$ and $v$, satisfy the scalar Helmholtz equation $\nabla^2 \phi + k^2 \phi = 0$ which elementary solutions in the spherical coordinate system are of the form:
\begin{subequations}\label{eq:anawaveeq_sol}
    \begin{align}
    \phi_{(l,n)} = &\cos{(l \varphi)} P_{n}^{l}(\cos{\theta}) z_{n}(kr), \\
    \phi_{(l,n)} = &\sin{(l \varphi)} P_{n}^{l}(\cos{\theta}) z_{n}(kr).
    \end{align}
\end{subequations}
In Eq.~(\ref{eq:anawaveeq_sol}), $l$ and $n$ are integers with $n \ge l \ge 0$, $P_{n}^{l}(\cos{\theta})$ is the associated Legendre polynomial, and $z_{n}(kr)$ is the spherical Bessel function. The following rules are applied to determine the choice of the function $z_{n}(kr)$. In the bounded domain comprising the origin, $z_{n}(kr) \equiv j_{n}(kr)$, the spherical Bessel function of the first kind, is used since $j_{n}(kr)$ is finite at the origin. In the unbounded external domain, the spherical Hankel function $z_{n}(kr) \equiv h_{n}(kr) = j_{n}(kr) + \rmi y_n(kr)$ is used since $\rmi k h_{n}(kr) \sim \rmi^{n} \exp{(\rmi k r)}/r$, represents an outgoing wave, where $y_{n}(kr)$ is the spherical Bessel function of the second kind. 

It is worth noting that the potentials $u$ and $v$ correspond to $\cos(l\varphi)$ and $\sin(l\varphi)$ formulations in Eq.~(\ref{eq:anawaveeq_sol}), respectively. Nevertheless, according to Eq.~(\ref{eq:EM_dip_sphcoor}), the fields driven by a concentric dipole in a sphere do not depend on $\varphi$. As such, only terms with $l=0$ in Eq.~(\ref{eq:anawaveeq_sol}) exist, which means just one potential is needed. Let us use the potential $u$ here. Moreover, when comparing the components given in Eq.~(\ref{eq:MNu}) and Eq.~(\ref{eq:EM_dip_sphcoor}), we can see that only the terms with $n=1$ in Eq.~(\ref{eq:anawaveeq_sol}) are needed. As a result, to obtain the induced field inside the sphere and the radiation field outside the sphere, we let
\begin{subequations}\label{eq:em_dip_u}
    \begin{align}
        u^{\rmout} &= C^{(1)}_{(0,1)} \, h_{1}^{(1)}(k_1 r) \cos{\theta},\\
        u^{\rmin} &= C^{(2)}_{(0,1)} \, j_{1}(k_2 r) \cos{\theta},
    \end{align}
\end{subequations}
where 
\begin{subequations}
    \begin{align}
        h_{1}^{(1)}(k_1 r) &= \frac{\sin(k_1 r)}{(k_1 r)^2} - \frac{\cos(k_1 r)}{k_1 r} + \rmi \left[-\frac{\cos(k_1 r)}{(k_1 r)^2} - \frac{\sin(k_1 r)}{k_1 r}\right], \\
        j_1(k_2 r) &= \frac{\sin(k_2 r)}{(k_2 r)^2} - \frac{\cos(k_2 r)}{k_2 r},
    \end{align}
\end{subequations}
and $C^{(1)}_{(0,1)}$ and $C^{(2)}_{(0,1)}$ are the coefficients to be determined by the boundary conditions.

Introducing Eq.~(\ref{eq:em_dip_u}) into Eq.~(\ref{eq:em_EH_MN}) and using Eq.~(\ref{eq:MNu}), we obtain
\begin{subequations}\label{eq:em_dip_esctr}
    \begin{align}
        E_{r}^{\rmout} &= - \rmi \left[\frac{1}{k_1} \frac{\partial^2(ru^{\rmout})}{\partial r^2} + k_1  r u^{\rmout}\right] = -C^{(1)}_{(0,1)} \frac{2\rmi}{k_1 r}h_1^{(1)}(k_1 r)  \cos{\theta}, \\
        E_{\theta}^{\rmout} &= -\frac{\rmi}{k_1 r} \frac{\partial^2(ru^{\rmout})}{\partial r \partial \theta} = C^{(1)}_{(0,1)} \frac{\rmi}{k_1 r}\left[ 2 h_1^{(1)}(k_1r) - k_1 r h_2^{(2)}(k_1 r) \right] \sin{\theta}, \\
        E_{\varphi}^{\rmout} & = 0, \\
        E_{r}^{\rmin} &= - \rmi \left[\frac{1}{k_2} \frac{\partial^2(ru^{\rmin})}{\partial r^2} + k_2  r u^{\rmin}\right] = -C^{(2)}_{(0,1)} \frac{2\rmi}{k_2 r}j_1(k_2 r)  \cos{\theta}, \\
        E_{\theta}^{\rmin} &= -\frac{\rmi}{k_1 r} \frac{\partial^2(ru^{\rmin})}{\partial r \partial \theta} = C^{(2)}_{(0,1)} \frac{\rmi}{k_2 r}\left[ 2 j_1(k_2 r) - k_2 r j_2(k_2 r) \right] \sin{\theta}, \\
        E_{\varphi}^{\rmin} & = 0;
    \end{align}
\end{subequations}
and 
\begin{subequations}\label{eq:em_dip_hsctr}
    \begin{align}
        H_{r}^{\rmout} &= 0, \\
        H_{\theta}^{\rmout} &= 0,\\
        H_{\varphi}^{\rmout} & = \sqrt{\frac{\epsilon_0\epsilon_1}{\mu_0 \mu_1}} \frac{1}{r} \frac{\partial (r u^{\rmout})}{\partial \theta} = -\sqrt{\frac{\epsilon_0\epsilon_1}{\mu_0 \mu_1}} C^{(1)}_{(0,1)} h_1^{(1)}(k_1 r) \sin{\theta}, \\
        H_{r}^{\rmin} &= 0, \\
        H_{\theta}^{\rmin} &= 0, \\
        H_{\varphi}^{\rmin} & = \sqrt{\frac{\epsilon_0\epsilon_2}{\mu_0 \mu_2}} \frac{1}{r} \frac{\partial (r u^{\rmin})}{\partial \theta} = -\sqrt{\frac{\epsilon_0\epsilon_2}{\mu_0 \mu_2}} C^{(2)}_{(0,1)} j_1(k_2 r) \sin{\theta}.
    \end{align}
\end{subequations}
The boundary conditions across the sphere surface at $r=a$ are:
\begin{subequations}
    \begin{align}
        \bs{E}^{\rmout}_{\parallel} &= \bs{E}^{\rmin}_{\parallel} + \bs{E}^{\rmEd}_{\parallel}, \\
        \bs{H}^{\rmout}_{\parallel} &= \bs{H}^{\rmin}_{\parallel} + \bs{H}^{\rmEd}_{\parallel},
    \end{align}
\end{subequations}
which means $E_{\theta}^{\rmout} = E_{\theta}^{\rmin} + E_{\theta}^{\rmEd}$ and $H_{\varphi}^{\rmout} = H_{\varphi}^{\rmin} + H_{\varphi}^{\rmEd}$. As such, using the relative formulations in Eqs.~(\ref{eq:EM_dip_sphcoor}), (\ref{eq:em_dip_esctr}) and (\ref{eq:em_dip_hsctr}), we have
\begin{subequations}\label{eq:anamatrix_C1C2}
    \begin{align}
        & C^{(1)}_{(0,1)} \frac{\rmi}{k_1 a}\left[ 2 h_1^{(1)}(k_1 a) - k_1 a h_2^{(2)}(k_1 a) \right] - C^{(2)}_{(0,1)} \frac{\rmi}{k_2 a}\left[ 2 j_1(k_2 a) - k_2 a j_2(k_2 a) \right] \nonumber \\
        = & \frac{1}{4\pi\epsilon_0\epsilon_2} p \left(\frac{1}{a^3} - \frac{\rmi k_2}{a^2} - \frac{k_2^2}{a} \right)\exp{(\rmi k_2a)}, \\
        & -\sqrt{\frac{\epsilon_0\epsilon_1}{\mu_0 \mu_1}} C^{(1)}_{(0,1)} h_1^{(1)}(k_1 a) + \sqrt{\frac{\epsilon_0\epsilon_2}{\mu_0 \mu_2}} C^{(2)}_{(0,1)} j_1(k_2 a) \nonumber \\
        =&  \frac{k_2^3}{4\pi \epsilon_0\epsilon_2 \mu_0 \mu_2 \omega} p \left( \frac{1}{\rmi k_2 a^2} - \frac{1}{a} \right)\exp{(\rmi k_2 a)}.
    \end{align}
\end{subequations}
Once $C^{(1)}_{(0,1)}$ and $C^{(2)}_{(0,1)}$ are obtained by solving Eq.~(\ref{eq:anamatrix_C1C2}), the electric and magnetic fields inside, outside and on the sphere surface can be obtained.

\medskip

%
\bibliographystyle{MSP}
\bibliography{ref}

\end{document}